\documentclass[aip,jcp,reprint,noshowkeys,superscriptaddress]{revtex4-2}
\usepackage{graphicx,dcolumn,bm,xcolor,microtype,multirow,amscd,amsmath,amssymb,amsfonts,physics,longtable,wrapfig,bbold,xspace,siunitx,soul}
\usepackage[version=4]{mhchem}

\usepackage[utf8]{inputenc}
\usepackage[T1]{fontenc}

\usepackage{hyperref}
\hypersetup{
    colorlinks,
    linkcolor={red!50!black},
    citecolor={red!70!black},
    urlcolor={red!80!black}
}

\usepackage{listings}
\definecolor{codegreen}{rgb}{0.58,0.4,0.2}
\definecolor{codegray}{rgb}{0.5,0.5,0.5}
\definecolor{codepurple}{rgb}{0.25,0.35,0.55}
\definecolor{codeblue}{rgb}{0.30,0.60,0.8}
\definecolor{backcolour}{rgb}{0.98,0.98,0.98}
\definecolor{mygray}{rgb}{0.5,0.5,0.5}

\definecolor{sqred}{rgb}{0.85,0.1,0.1}
\definecolor{sqgreen}{rgb}{0.25,0.65,0.15}
\definecolor{sqorange}{rgb}{0.90,0.50,0.15}
\definecolor{sqblue}{rgb}{0.10,0.3,0.60}

\lstdefinestyle{mystyle}{
    backgroundcolor=\color{backcolour},
    commentstyle=\color{codegreen},
    keywordstyle=\color{codeblue},
    numberstyle=\tiny\color{codegray},
    stringstyle=\color{codepurple},
    basicstyle=\ttfamily\footnotesize,
    breakatwhitespace=false,
    breaklines=true,
    captionpos=b,
    keepspaces=true,
    numbers=left,
    numbersep=5pt,
    numberstyle=\ttfamily\tiny\color{mygray},
    showspaces=false,
    showstringspaces=false,
    showtabs=false,
    tabsize=2
  }

  \newcolumntype{d}{D{.}{.}{-1}}

\newcommand{\mc}{\multicolumn}

\newcommand{\SupInf}{\textcolor{blue}{SI}}

\newcommand{\fk}[1]{\textcolor{black}{#1}}

\usepackage{comment}

\renewcommand{\tr}[2]{($#1\rightarrow#2$)}

\newcommand{\Fp}{$^a$}
\newcommand{\Fd}{$^a$}
\newcommand{\Ft}{$^a$}
\newcommand{\Qp}{$^b$}

\newcommand{\qp}{$^c$}
\newcommand{\qd}{$^c$}
\newcommand{\Tp}{$^d$}

\newcommand{\Cp}{$^e$}
\newcommand{\Cd}{$^e$}
\newcommand{\Ct}{$^e$}
\newcommand{\Np}{$^f$}
\newcommand{\Nd}{$^f$}
\newcommand{\Nt}{$^f$}

\newcommand{\qdb}{$^g$}
\newcommand{\Tdb}{$^h$}
\newcommand{\Cdb}{$^i$}
\newcommand{\Ndb}{$^j$}

\newcommand{\Qtb}{$^k$}
\newcommand{\qtb}{$^l$}
\newcommand{\Ttb}{$^m$}
\newcommand{\ttb}{$^n$}
\newcommand{\Ctb}{$^o$}
\newcommand{\Ntb}{$^p$}

\newcommand{\LCPQ}{Laboratoire de Chimie et Physique Quantiques (UMR 5626), Universit\'e de Toulouse, CNRS, UPS, France}
\newcommand{\CEISAM}{Nantes Universit\'e, CNRS,  CEISAM UMR 6230, F-44000 Nantes, France}
\newcommand{\IUF}{Institut Universitaire de France (IUF), F-75005 Paris, France}

\begin{document}

\title{Reference Energies for Double Excitations: Improvement and Extension}

\author{F\'abris \surname{Kossoski}}
	\email{fkossoski@irsamc.ups-tlse.fr}
	\affiliation{\LCPQ}
\author{Martial \surname{Boggio-Pasqua}}
	\email{martial.boggio@irsamc.ups-tlse.fr}
	\affiliation{\LCPQ}
\author{Pierre-Fran\c{c}ois \surname{Loos}}
	\email{loos@irsamc.ups-tlse.fr}
	\affiliation{\LCPQ}
\author{Denis \surname{Jacquemin}}
	\email{Denis.Jacquemin@univ-nantes.fr}
	\affiliation{\CEISAM}
	\affiliation{\IUF}

\begin{abstract}
In the realm of photochemistry, the significance of double excitations (also known as doubly-excited states), where two electrons are concurrently elevated to higher energy levels, lies in their involvement in key electronic transitions essential in light-induced chemical reactions as well as their challenging nature from the computational theoretical chemistry point of view.
Based on state-of-the-art electronic structure methods (such as high-order coupled-cluster, selected configuration interaction, and multiconfigurational methods), we improve and expand our prior set of accurate reference excitation energies for electronic states exhibiting a substantial amount of double excitations [\href{http://dx.doi.org/10.1021/acs.jctc.8b01205}{Loos et al. \textit{J. Chem. Theory Comput.} \textbf{2019}, \textit{15}, 1939}]. This extended collection encompasses 47 electronic transitions across 26 molecular systems that we separate into two distinct subsets:
(i) 28 ``genuine'' doubly-excited states where the transitions almost exclusively involve doubly-excited configurations and
(ii) 19 ``partial'' doubly-excited states which exhibit a more balanced character between singly- and doubly-excited configurations.
For each subset, we assess the performance of high-order coupled-cluster (CC3, CCSDT, CC4, and CCSDTQ) and multiconfigurational methods (CASPT2, CASPT3, PC-NEVPT2, and SC-NEVPT2).
Using as a probe the percentage of single excitations involved in a given transition ($\%T_1$) computed at the CC3 level, we also propose a simple correction that reduces the errors of CC3 by a factor of 3, for both sets of excitations.
We hope that this more complete and diverse compilation of double excitations will help future developments of electronic excited-state methodologies.
\bigskip
\begin{center}
  \boxed{\includegraphics[width=0.4\textwidth]{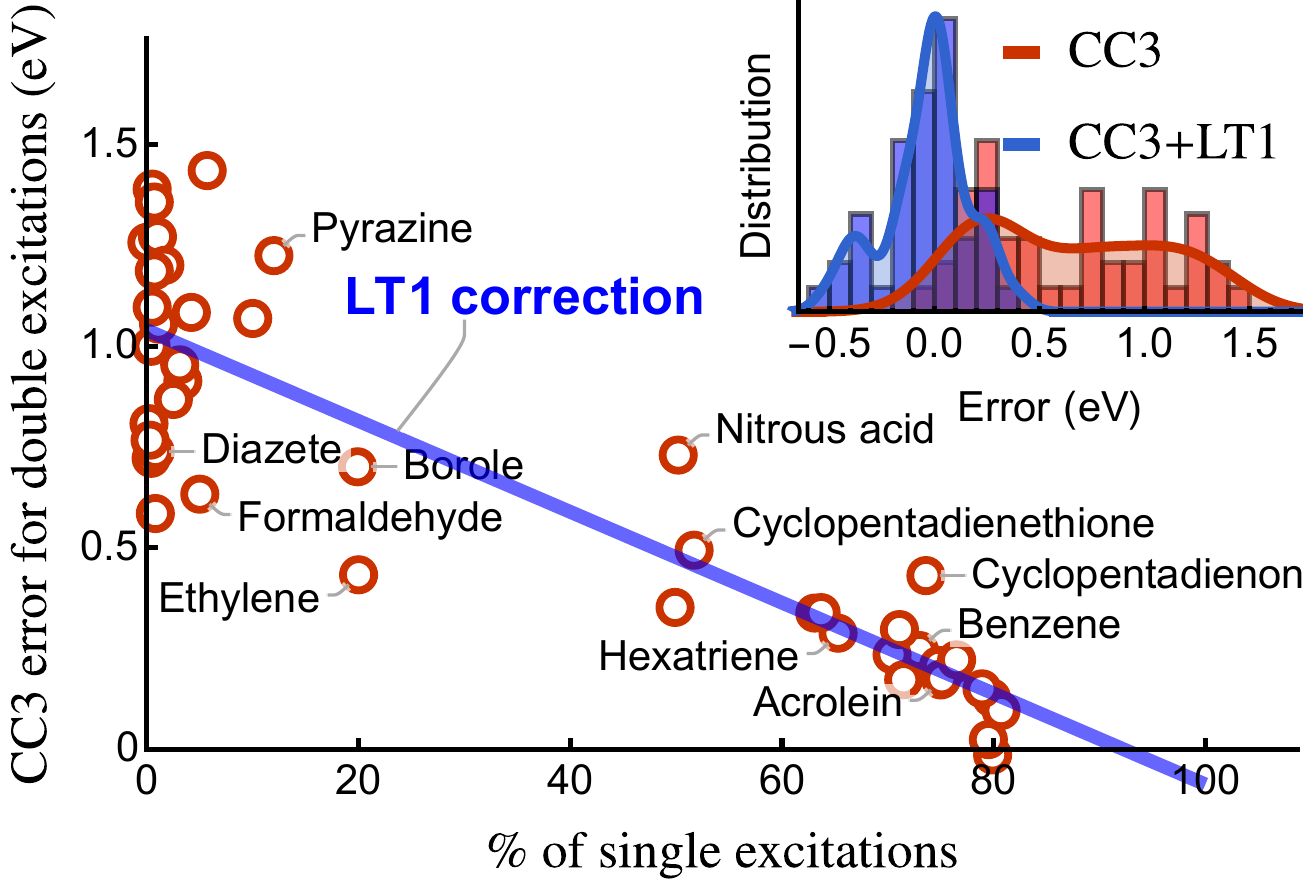}}
\end{center}
\end{abstract}
\maketitle

\section{Introduction}
\label{sec:introduction}
When discussing optical excitations in molecules, it is often useful to specify the number of electrons involved in a particular excitation. In the simple orbital picture, a ``single'' excitation refers to the promotion of a single electron from an occupied orbital, or hole (h) state, to a vacant orbital, or particle (p) state, resulting in a ``singly-excited'' (or 1h1p) configuration. In contrast, a ``double'' excitation involves the promotion of a pair of electrons with opposite or same spins from the same or different occupied orbitals to vacant ones, leading to a ``doubly-excited'' (or 2h2p) configuration.

In the framework of many-body perturbation theory, singly-excited states correspond to bound states between an optically excited electron and the hole left behind due to the excitation. This quasiparticle state is known as an exciton. On the other hand, doubly-excited states correspond to the appearance of a biexciton, or a pair of excitons, which consists of two holes and two excited electrons bound together. At the many-body level, biexcitons involve four-body correlations, which explains the difficulty behind their theoretical treatment, whereas experiments also struggle to characterize them.

These multiple excited states can serve as rapid relaxation pathways after the absorption of highly energized photons to prevent damage \cite{Abramavicius_2008} or utilize the extra energy deposited in nanocrystal quantum dots by highly energetic photons. \cite{Klimov_2000} These multiple exciton states exhibit correlations between their constituent electrons and holes and can be observed in non-linear multidimensional optical spectroscopy. \cite{Stone_2009,Biswas_2022} Direct spectroscopic investigation of biexcitons is challenging because the optical transitions are formally forbidden, hence their classification as ``dark'' states.
However, these states can be indirectly probed through techniques like photoluminescence, \cite{Miller_1982} spectrally resolved four-wave mixing experiments, \cite{Feuerbacher_1991} \fk{or excited-state absorption. \cite{Mikhailov_2008,Fischer_2015}}

Because this assignment is only meaningful within a given wave function method and set of reference orbitals, the concept of single or double excitations is not precisely defined, and it is more accurate to refer to double excitations as \textit{``electronic states with doubly-excited character''}.
Classifying doubly-excited states is a tedious task, \cite{doCasal_2023} that has sparked controversies in recent years, particularly concerning the lowest dark state of butadiene. \cite{Shu_2017,Barca_2018b,doCasal_2023,Kossoski_2023}

Doubly-excited states are ubiquitous in many photochemical mechanisms, such as the photochemistry of polyenes, \cite{Serrano-Andres_1993,Cave_1988b,Lappe_2000,Boggio-Pasqua_2004,Maitra_2004,Cave_2004,Wanko_2005,Starcke_2006,Catalan_2006,Mazur_2009,Angeli_2010,Mazur_2011,Huix-Rotllant_2011} in singlet fission \cite{Smith_2010} where they can play a major role in the formation of the triplet-pair state, \cite{SandovalSalinas_2020} in thermally activated delayed fluorescence (TADF) \cite{Endo_2009,Uoyama_2012,Di-Maiolo_2024} where their predominant interaction with singlet states can produce molecules with an inverted singlet-triplet gap, \cite{deSilva_2019,Sanz-Rodrigo_2021,Ricci_2021,Olivier_2017,Olivier_2018,SanchoGarca_2022,Curtis_2023,Loos_2023} i.e., where the lowest singlet excited state is lower in energy than the lowest triplet state.

Double excitations pose significant challenges for adiabatic time-dependent density-functional theory (TD-DFT), \cite{Runge_1984,Petersilka_1996,Casida,UlrichBook} which only considers explicitly the 1h1p configurations. To capture doubly-excited states, one must go beyond the adiabatic approximation which is a non-trivial task.\cite{Levine_2006,Tozer_2000,Elliott_2011,Loos_2019a,Maitra_2004,Cave_2004,Mazur_2009,Mazur_2011,Huix-Rotllant_2011,Elliott_2011,Maitra_2012} The Bethe-Salpeter equation (BSE) formalism \cite{Salpeter_1951,Strinati_1988,Blase_2018,Blase_2020} encounters a similar challenge, requiring frequency-dependent kernels to compute doubly-excited states. \cite{Strinati_1988,Rohlfing_2000,Romaniello_2009a,Sangalli_2011,Loos_2020h,Authier_2020,Bintrim_2022,Monino_2023} This is why more advanced theories should be used.
\fk{Second-order wave function methods like CIS(D), \cite{Head-Gordon_1994,Head-Gordon_1995} EOM-CC2, \cite{Christiansen_1995a,Hattig_2000} and ADC(2) \cite{Trofimov_1997,Dreuw_2015,Schirmer_2018} do not explicitly account for the 2h2p configurations.
Even though they are fully described in EOM-CCSD, \cite{Purvis_1982,Scuseria_1987,Koch_1990a,Koch_1990b,Stanton_1993a,Stanton_1993b} this method still provides insufficient accuracy for double excitations.}
In short, the most computationally effective theories are inadequate to tackle doubly-excited states.

Although ADC(3) includes the 2h2p configurations, \cite{Trofimov_2002,Harbach_2014,Dreuw_2015,Schirmer_2018} it struggles to deliver accurate transition energies for doubly-excited states. \cite{Loos_2020d} A recent study have demonstrated that ADC(4), which additionally incorporates explicitly the 3h3p configurations, is better suited for this purpose. \cite{Leitner_2022}

Spin-flip methods offer an alternative approach where ground and doubly-excited states are accessed through single excitations from the lowest triplet state. \cite{Huix-Rotllant_2010,Krylov_2001b,Shao_2003,Wang_2004,Wang_2006,Minezawa_2009,Monino_2021} Pushing the excitation degree to triple or quadruple excitations within coupled-cluster (CC) theory \cite{Cizek_1966,Cizek_1969,Paldus_1992,Crawford_2000,Bartlett_2007,Shavitt_2009} also enables to capture double excitations. \cite{Watson_2012,Loos_2018a,Loos_2019c,Loos_2020c}

Multiconfigurational self-consistent field methods, such as complete-active-space self-consistent field (CASSCF), and their second-order perturbatively corrected variants, CASPT2 \cite{Andersson_1990,Andersson_1992,Roos_1995a,Battaglia_2023} and NEVPT2, \cite{Angeli_2001a,Angeli_2001b,Angeli_2002,Angeli_2006} naturally handle single and double excitations on an equal footing. \cite{Schapiro_2013,Sarkar_2022,Boggio-Pasqua_2022}

All of these methods have well-documented strengths and weaknesses, and it is beneficial to utilize each of them in their optimal scenarios, capitalizing on their respective advantages. For example, CASPT2 and NEVPT2 suffer from exponential growth in computational cost with the number of active electrons and orbitals. Careful selection of the active space can provide access to doubly-excited states at a manageable expense, albeit with limitations imposed by the active space choice. In practical applications, accuracy in the 0.1--0.2 eV range can be achieved by employing appropriate active spaces.

EOM-CC methods offer a systematically improvable pathway to high accuracy, exemplified by the series of methods EOM-CC3, \cite{Christiansen_1995b,Koch_1995,Koch_1997,Hald_2001,Paul_2021} EOM-CCSDT, \cite{Noga_1987,Scuseria_1988,Watts_1994,Kucharski_2001} EOM-CC4, \cite{Kallay_2004b,Kallay_2005,Loos_2021a,Loos_2022a} and EOM-CCSDTQ. \cite{Kucharski_1991,Kallay_2001,Hirata_2004,Kallay_2003,Kallay_2004a} Although the cost increases with the maximum excitation degree, it remains polynomial. By including quadruples, chemical accuracy (i.e., an absolute error below \SI{0.043}{\eV}) can be reached for double excitations. \cite{Loos_2021c,Loos_2022b}

For small systems, one can also compute full CI (FCI) excitation energies that are usually obtained with selected CI (SCI) methods for computational efficiency.\cite{Giner_2013,Giner_2015,Holmes_2017,Mussard_2018,Tubman_2018,Chien_2018,Tubman_2020,Loos_2020i,Yao_2020,Damour_2021,Yao_2021,Larsson_2022,Coe_2022}
The \textit{``Configuration Interaction using a Perturbative Selection made Iteratively''} (CIPSI) algorithm \cite{Huron_1973,Evangelisti_1983,Cimiraglia_1985,Cimiraglia_1987,Illas_1988,Povill_1992,Giner_2013,Giner_2015,Garniron_2019} is particularly effective in computing vertical excitation energies in small molecules. \cite{Loos_2018a,Loos_2019c,Loos_2020c,Loos_2020f,Veril_2021,Loos_2021a,Loos_2021b,Garniron_2018,Giner_2019} Other SCI methods, such as semistochastic heatbath CI (SHCI), \cite{Holmes_2017,Chien_2018} Monte Carlo CI (MCCI), \cite{Prentice_2019,Coe_2022} iterative CI (iCI), \cite{Zhang_2020} and adaptive sampling CI (ASCI), \cite{Schriber_2017} have also been successfully employed for computing excited states.

In a 2019 paper, we presented reference energies for double excitations in a collection of 14 small- and medium-sized molecules, encompassing a set of 20 vertical transitions. \cite{Loos_2019c} For the majority of these calculations, we relied on FCI computations to establish our ``theoretical best estimates'' (TBEs). However, for the largest molecules in the dataset, we employed high-order EOM-CC or multiconfigurational methods to achieve accurate estimates.

These reference energies for doubly-excited states have proven to be valuable to the scientific community, serving as benchmarks and test cases for new computational methods, including orbital-optimized DFT, \cite{Hait_2020,Hait_2021,Zhang_2023} ensemble DFT, \cite{Gould_2022} quantum Monte Carlo, \cite{Otis_2020,Shepard_2022,Otis_2023} EOM-CC, \cite{Ravi_2022,Rishi_2023} and others, \cite{Leitner_2022,Kossoski_2023,Dombrowski_2023,Sarkar_2022,Boggio-Pasqua_2022,King_2022,Wang_2023} and have now been incorporated into \textsc{quest}, a comprehensive and diverse database of highly accurate vertical excitation energies for small- and medium-sized molecules. \cite{Loos_2020d,Veril_2021}

In this work, using the same computational protocol, \cite{Loos_2018a,Loos_2019c,Loos_2020c,Loos_2020f,Veril_2021,Loos_2021a,Loos_2021b} we broaden and enhance this set of reference data.
Firstly, we significantly augment our collection of double excitations by reporting 47 vertical excitation energies for electronic states exhibiting substantial and more balanced doubly-excited character.
We classify these states as ``genuine'' (where transition primarily involves doubly-excited determinants) and
``partial'' (where the transition involves a more balanced character between singly- and doubly-excited determinants).
Secondly, we refine the previous TBEs by incorporating new FCI, high-order CC, and multiconfigurational (NEVPT2, CASPT2, and CASPT3) calculations to improve their accuracy.
\fk{We stress that our TBEs concern vertical excitation energies. Since vibrational effects are not considered here, comparisons between the TBEs and experimental results should be made with caution. \cite{Loos_2019b}}

\section{Computational details}
\label{sec:comp}

The ground-state geometries are extracted from Ref.~\onlinecite{Loos_2019c} or optimized using the same protocol, that is, at the CC3/aug-cc-pVTZ level of theory for most systems. These geometries can be found in the supporting information ({\SupInf}). Calculations are performed for all systems using three Gaussian atomic basis sets: 6-31+G* \fk{(with spherical d functions)}, aug-cc-pVDZ, and aug-cc-pVTZ,
which are sometimes referred to as Pop, AVDZ, and AVTZ from here on. All excited-state calculations were performed using the frozen-core approach (except for the \ce{Be} atom).

Vertical excitation energies were obtained using FCI calculations employing the CIPSI algorithm. \cite{Huron_1973,Giner_2013,Giner_2015,Garniron_2017,Garniron_2018} All calculations are performed using \textsc{quantum package} \cite{Garniron_2019} following the same protocol as our previous studies. \cite{Loos_2018a,Loos_2019c,Loos_2020c}
Here, the extrapolated FCI estimate obtained with the CIPSI calculations is referred to as exFCI. \cite{Burton_2024}
Extrapolation errors are estimated following the procedure outlined in Ref.~\onlinecite{Veril_2021}.

Most CC calculations were conducted using \textsc{cfour}, \cite{Matthews_2020} which offers an efficient implementation of high-order CC methods up to quadruples. \cite{Matthews_2015} For CC3 calculations, we utilize \textsc{dalton}, \cite{dalton} while \textsc{mrcc} \cite{Kallay_2020} is employed for CC calculations that go beyond quadruples.
\fk{To obtain excitation energies, \textsc{cfour} relies on the EOM approach, whereas \textsc{dalton} and \textsc{mrcc} employ the linear response formalism. Both frameworks yield identical excitation energies. \cite{Rowe_1968a,Stanton_1993a}}

Multiconfigurational calculations are performed using \textsc{molpro} \cite{Werner_2020} following the protocol described in Ref.~\onlinecite{Loos_2019c}.
The {\SupInf} provides additional details and a description of the active spaces for each symmetry representation for each system and transition.
We performed NEVPT2 calculations in the partially contracted (PC) and in the strongly-contracted (SC) schemes,
as well as CASPT2 and CASPT3 calculations, \cite{Werner_1996,Boggio-Pasqua_2022} both with and without an IPEA shift. \cite{Andersson_1993,Andersson_1995,Ghigo_2004}
To differentiate the two cases, throughout the text, we indicate in parenthesis when the IPEA shift is adopted: CASPT2(IPEA) uses it and CASPT2 does not. The IPEA shift is set to its default value of \SI{0.25}{\hartree}.
Unless otherwise stated, all CASPT2 and CASPT3 calculations have been performed with a level shift of $0.3$ a.u. Note that the Fock operator used in the zeroth-order Hamiltonian was systematically computed using the state-specific density.

To obtain a basic definition of the singly- or doubly-excited character, we use the $\%T_1$ diagnostic computed at the CC3/AVTZ level, \fk{as implemented in \textsc{dalton}, where one sums the weights of all singly-excited response vectors.}
A ``genuine'' double excitation corresponds to \fk{$\%T_1 \lesssim 20\%$},
whereas a ``partial'' double excitation, where the transition is dominated by single excitations, has intermediate values, up to $\%T_1 \approx 85\%$.
\fk{The few cases where $\%T_1 \approx 50\%$ are discussed separately.}
Transitions characterized by $\%T_1 \gtrsim 85\%$ correspond to singly-excited states and are not addressed here.
With few exceptions (discussed below), the classification based on the $\%T_1$ diagnostic computed at the CC3 level agrees
with that based on inspection of the CIPSI wave function or the amplitudes in high-order CC calculations.

\begin{figure*}
\includegraphics[width=\linewidth,viewport=1cm 2.5cm 27cm 20.5cm, clip]{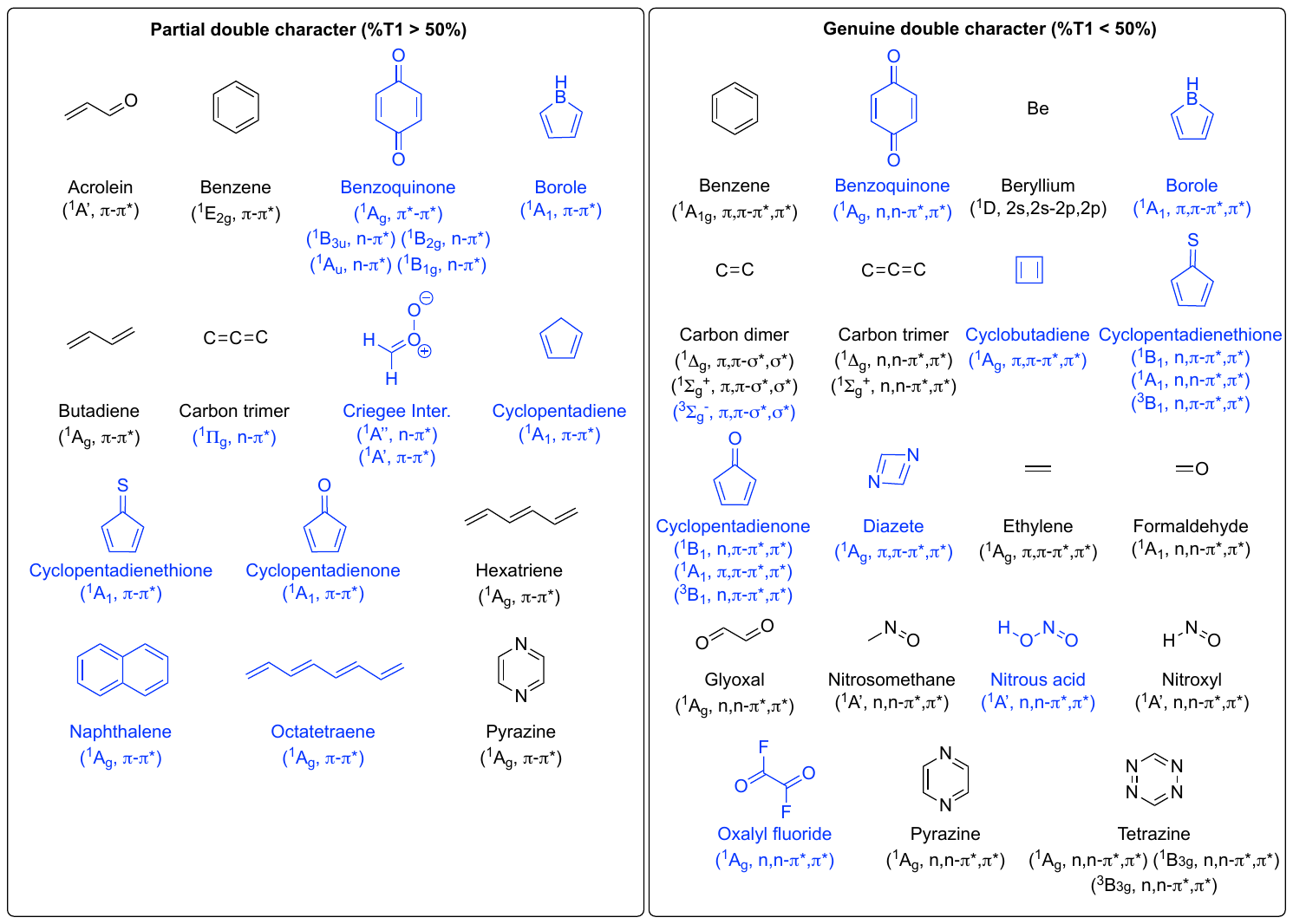}
\caption{List of molecules considered in the present study having states with partial double excitation character (left) and genuine double excitation character (right). In comparison to Ref.~\onlinecite{Loos_2019c}, the additional systems and states are highlighted in blue.}
\label{fig:mol}
\end{figure*}

Here we address 47 excited states having doubly-excited character, encompassing the 26 molecules depicted in Fig.~\ref{fig:mol}.
We have classified 19 as ``partial'' double excitations, 14 of them are new compared to Ref.~\onlinecite{Loos_2019c} and we have performed additional calculations for the others.
The other 28 states were classified as ``genuine'' double excitations,
13 of them are new in comparison to Ref.~\onlinecite{Loos_2019c}, but we also performed new calculations for the others.
In comparison to Ref.~\onlinecite{Loos_2019c}, the present set thus substantially increases the pool of states possessing doubly-excited character, from 15 to 28 genuine double excitations,
and, most noticeably, from 5 to 19 partial double excitations.
This allows us to draw more general conclusions about the performance of various methodologies in describing doubly-excited states.
In line with the latest contributions to the \textsc{quest} database, \cite{Jacquemin_2023,Loos_2023,Loos_2024,Marie_2024} all excitation energies reported here are given with three decimal places.

We provide a TBE for each basis set (Pop, AVDZ, and AVTZ). The result for the largest one, TBE/AVTZ, should be taken as our most accurate estimate of the excitation energy.
For the sake of conciseness, the TBE/AVTZ is referred to as TBE from hereon, whereas the smaller ones are explicitly stated (TBE/Pop and TBE/AVDZ).
When a given TBE is chemically accurate (estimated uncertainty less than \SI{1}{kcal/mol} or \SI{0.043}{\eV}), we label it as safe.
Otherwise, we referred to the TBE as unsafe. For borderline cases, we conservatively assign them as unsafe.

Starting with the smallest 6-31+G* basis set, the most accurate level of theory we are able to employ yields the TBE/Pop.
When increasing the basis set, the same level of theory is often impractical, and we adopt instead less demanding and more approximate methods.
This happens in the majority of cases, where we resort to composite models to compute the TBE.
The result from a smaller basis set and accurate level of theory, say CCSDTQ in the 6-31+G* basis set (CCSDTQ/Pop),
is combined with that from a larger basis set and less accurate level of theory, say CC4 in the AVDZ basis set (CC4/AVDZ),
to yield a TBE/AVDZ $=$ CCSDTQ/Pop $+$ [CC4/AVDZ $-$ CC4/Pop].
Here we refer to the term in brackets as a basis set correction.
We often employ two different levels of theory for each basis set correction.
Following on the previous example, CCSDT can be used to provide the AVTZ basis set correction,
such that TBE $=$ TBE/AVDZ $+$ [CCSDT/AVTZ $-$ CCSDT/AVDZ].

\fk{It is not always straightforward to choose which family of methods (CIPSI, CC, or multiconfigurational) yields the most accurate excitation energy for a given basis set,
and, similarly, which method is more trustworthy for the basis set correction.
Although increasing the excitation degree in CC yields more accurate results, it is not obvious how the accuracy of a given CC method compares with that of a given multiconfigurational method.
Likewise, CIPSI calculations systematically approach the exact result but has an associated extrapolation error, making its comparison with CC or multiconfigurational methods less evident.}
This is further complicated by the fact that the performance of the methods can significantly depend on whether the doubly-excited state has a genuine or partial character.
Here we adopted the following guiding principles.
When CIPSI calculations yield an extrapolation error smaller than the chemical accuracy, this result is taken as a safe TBE for that basis set.
We achieved this level of convergence in CIPSI calculations only for relatively small systems (such as carbon trimer in the AVTZ basis)
and/or basis set (for instance, the $^1B_1$ state of cyclopentadienone in the 6-31+G* basis).
When larger extrapolation errors are obtained, the upper and lower bounds of the CIPSI result can guide the choice of a higher-order CC or multiconfigurational method for the TBE.
When available, CCSDTQ or CC4 is preferred over multiconfigurational methods, which is justified \emph{a posteriori} by their overall superior performance, as discussed below.
CCSDT remains more accurate than multiconfigurational alternatives for partial double excitations, while the opposite is found for genuine double excitations, which is also justified \emph{a posteriori}.
In the latter case, selecting the most reliable multiconfigurational method is less trivial.
The extrapolation error of CIPSI, the very likely overestimated energies produced by CC, and the global statistics,
all provide valuable bounds and insights to determine which method is the most reliable to produce the TBE.
In light of the present results, we further elaborate on the choice of method for the basis set correction in Sec.~\ref{sec:stat}.

\section{Double Excitations: Improvement}
\label{sec:Improvement}

Table~\ref{tab:energies} contains the vertical excitation energies for the 47 excited states considered here,
as obtained with various levels of theory and basis sets,
along with the TBEs for each basis and the methods employed to obtain the TBEs.
The TBEs are gathered in Table~\ref{tab:tbe},
along with a comparison with previously reported TBEs. \cite{Loos_2019c,Veril_2021,Loos_2022b,Monino_2022a}

\clearpage

\begin{squeezetable}
\begin{longtable}{llll}
\caption{Vertical excitation energies (in \si{\eV}) computed at various levels of theory, and for the 6-31+G*, aug-cc-pVDZ, and aug-cc-pVTZ basis sets (abbreviated to Pop, AVDZ, and AVTZ)
along with the TBEs for each basis and the methods employed to obtain the TBEs. Each state is labeled by its spatial and spin symmetries and by its $\%T_1$ value obtained at the CC3/AVTZ level of theory.}
\label{tab:energies}
\\
\hline\hline
& \mc{3}{c}{Basis Set} \\
\cline{2-4}
& \mc{1}{c}{6-31+G*} & \mc{1}{c}{aug-cc-pVDZ} & \mc{1}{c}{aug-cc-pVTZ} \\
\hline
\endfirsthead
\hline\hline
& \mc{3}{c}{Basis Set} \\
\cline{2-4}
& \mc{1}{c}{6-31+G*} & \mc{1}{c}{aug-cc-pVDZ} & \mc{1}{c}{aug-cc-pVTZ} \\
\hline
\endhead
\hline \multicolumn{4}{r}{Continued on next page} \\
\endfoot
\hline\hline
\multicolumn{4}{l}{\Fp exFCI.} \\
\multicolumn{4}{l}{\Qp CCSDTQ.} \\
\multicolumn{4}{l}{\qp CC4.} \\
\multicolumn{4}{l}{\Tp CCSDT.} \\
\multicolumn{4}{l}{\Cp CASPT3.} \\
\multicolumn{4}{l}{\Np PC-NEVPT2.} \\
\multicolumn{4}{l}{\qdb TBE/Pop + CC4/AVDZ $-$ CC4/Pop.} \\
\multicolumn{4}{l}{\Tdb TBE/Pop + CCSDT/AVDZ $-$ CCSDT/Pop.} \\
\multicolumn{4}{l}{\Cdb TBE/Pop + CASPT3/AVDZ $-$ CASPT3/Pop.} \\
\multicolumn{4}{l}{\Ndb TBE/Pop + PC-NEVPT2/AVDZ $-$ PC-NEVPT2/Pop.} \\
\multicolumn{4}{l}{\Qtb TBE/AVDZ + CCSDTQ/AVTZ $-$ CCSDTQ/AVDZ.} \\
\multicolumn{4}{l}{\qtb TBE/AVDZ + CC4/AVTZ $-$ CC4/AVDZ.} \\
\multicolumn{4}{l}{\Ttb TBE/AVDZ + CCSDT/AVTZ $-$ CCSDT/AVDZ.} \\
\multicolumn{4}{l}{\ttb TBE/AVDZ + CC3/AVTZ $-$ CC3/AVDZ.} \\
\multicolumn{4}{l}{\Ctb TBE/AVDZ + CASPT3/AVTZ $-$ CASPT3/AVDZ.} \\
\multicolumn{4}{l}{\Ntb TBE/AVDZ + PC-NEVPT2/AVTZ $-$ PC-NEVPT2/AVDZ.} \\
\endlastfoot
\mc{4}{c}{Acrolein, $^1A'$, $\%T_1 = 75\%$} \\
                        TBE              & 8.034\Qp & 7.940\qdb & 7.928\Ttb \\
                        FCI              & 8.00(3)&        &        \\
                        CCSDTQ           & 8.034  &        &        \\
                        CC4              & 8.035  & 7.941  &        \\
                        CCSDT            & 8.110  & 8.024  & 8.012  \\
                        CC3              & 8.206  & 8.106  & 8.078  \\
                        CASPT2           & 7.634  & 7.626  & 7.528  \\
                        CASPT2(IPEA)     & 8.028  & 8.036  & 7.961  \\
                        CASPT3           & 7.976  & 7.969  & 7.906  \\
                        CASPT3(IPEA)     & 8.048  & 8.046  & 7.979  \\
                        SC-NEVPT2        & 8.082  & 8.085  & 8.006  \\
                        PC-NEVPT2        & 7.909  & 7.930  & 7.846  \\
\hline
\mc{4}{c}{Benzene, $^1E_{2g}$, $\%T_1 = 73.\%$} \\
                        TBE              & 8.259\qp & 8.211\qd & 8.190\Ttb \\
                        CC4              & 8.259  & 8.211  &       \\
                        CCSDT            & 8.424  & 8.380  & 8.359 \\
                        CC3              & 8.504  & 8.439  & 8.381 \\
                        CASPT2           & 7.986  & 7.901  & 7.816 \\
                        CASPT2(IPEA)     & 8.408  & 8.374  & 8.314 \\
                        CASPT3           & 8.251  & 8.220  & 8.162 \\
                        CASPT3(IPEA)     & 8.342  & 8.325  & 8.263 \\
                        SC-NEVPT2        & 8.623  & 8.606  & 8.555 \\
                        PC-NEVPT2        & 8.584  & 8.564  & 8.512 \\
\hline
\mc{4}{c}{Benzene, $^1A_{1g}$} \\
                        TBE              & 10.531\Cp & 10.360\Cd & 10.315\Ct \\
                        CASPT2           &  9.698 &  9.463 &  9.326 \\
                        CASPT2(IPEA)     & 10.501 & 10.343 & 10.236 \\
                        CASPT3           & 10.531 & 10.360 & 10.315 \\
                        CASPT3(IPEA)     & 10.676 & 10.528 & 10.468 \\
                        SC-NEVPT2        & 10.633 & 10.479 & 10.381 \\
                        PC-NEVPT2        & 10.349 & 10.179 &  9.995 \\
\hline
\mc{4}{c}{Benzoquinone, $^1A_g$, $\%T_1 = 63\%$} \\
                        TBE              & 6.339\qp & 6.381\Tdb & 6.351\ttb \\
                        CC4              & 6.339  &        &       \\
                        CCSDT            & 6.600  & 6.642  &       \\
                        CC3              & 6.677  & 6.701  & 6.671 \\
                        CASPT2           & 5.960  & 5.934  & 5.849 \\
                        CASPT2(IPEA)     & 6.341  & 6.354  & 6.287 \\
                        CASPT3           & 6.355  & 6.348  & 6.303 \\
                        CASPT3(IPEA)     & 6.390  & 6.394  & 6.343 \\
                        SC-NEVPT2        & 6.471  & 6.492  & 6.431 \\
                        PC-NEVPT2        & 6.435  & 6.452  & 6.390 \\
\hline
\mc{4}{c}{Benzoquinone, $^1B_{3u}$, $\%T_1 = 80\%$} \\
                        TBE              & 5.819\qp & 5.670\Tdb & 5.656\ttb \\
                        CC4              & 5.819  &        &       \\
                        CCSDT            & 5.959  & 5.810  &       \\
                        CC3              & 5.947  & 5.769  & 5.755 \\
                        CASPT2           & 5.211  & 5.100  & 5.002 \\
                        CASPT2(IPEA)     & 5.895  & 5.841  & 5.774 \\
                        CASPT3           & 6.051  & 5.985  & 5.954 \\
                        CASPT3(IPEA)     & 6.121  & 6.073  & 6.036 \\
                        SC-NEVPT2        & 6.114  & 6.074  & 6.020 \\
                        PC-NEVPT2        & 6.048  & 5.999  & 5.944 \\
\hline
\mc{4}{c}{Benzoquinone, $^1B_{2g}$, $\%T_1 = 76\%$} \\
                        TBE              & 5.898\qp & 5.766\Tdb & 5.764\ttb \\
                        CC4              & 5.898  &        &       \\
                        CCSDT            & 6.104  & 5.972  &       \\
                        CC3              & 6.108  & 5.937  & 5.935 \\
                        CASPT2           & 5.213  & 5.102  & 5.005 \\
                        CASPT2(IPEA)     & 5.882  & 5.828  & 5.761 \\
                        CASPT3           & 6.046  & 5.982  & 5.950 \\
                        CASPT3(IPEA)     & 6.113  & 6.066  & 6.028 \\
                        SC-NEVPT2        & 6.099  & 6.059  & 6.004 \\
                        PC-NEVPT2        & 6.037  & 5.988  & 5.932 \\
\hline
\mc{4}{c}{Benzoquinone, $^1A_u$, $\%T_1 = 75\%$} \\
                        TBE              & 6.200\qp & 6.109\Tdb & 6.083\ttb \\
                        CC4              & 6.200  &        &       \\
                        CCSDT            & 6.463  & 6.372  &       \\
                        CC3              & 6.408  & 6.295  & 6.269 \\
                        CASPT2           & 5.742  & 5.650  & 5.547 \\
                        CASPT2(IPEA)     & 6.315  & 6.281  & 6.210 \\
                        CASPT3           & 6.462  & 6.402  & 6.352 \\
                        CASPT3(IPEA)     & 6.521  & 6.477  & 6.422 \\
                        SC-NEVPT2        & 6.617  & 6.603  & 6.548 \\
                        PC-NEVPT2        & 6.581  & 6.558  & 6.502 \\
\hline
\mc{4}{c}{Benzoquinone, $^1B_{1g}$, $\%T_1 = 70\%$} \\
                        TBE              & 6.529\qp & 6.495\Tdb & 6.469\ttb \\
                        CC4              & 6.529  &        &       \\
                        CCSDT            & 6.773  & 6.739  &       \\
                        CC3              & 6.764  & 6.707  & 6.681 \\
                        CASPT2           & 5.693  & 5.596  & 5.479 \\
                        CASPT2(IPEA)     & 6.336  & 6.301  & 6.234 \\
                        CASPT3           & 6.581  & 6.523  & 6.492 \\
                        CASPT3(IPEA)     & 6.547  & 6.503  & 6.451 \\
                        SC-NEVPT2        & 6.641  & 6.625  & 6.574 \\
                        PC-NEVPT2        & 6.606  & 6.581  & 6.530 \\
\hline
\mc{4}{c}{Benzoquinone, $^1A_g$, $\%T_1 = 0\%$} \\
                        TBE              & 4.665\Np & 4.593\Nd & 4.566\Nt \\
                        CC4              & 4.771  &        &       \\
                        CCSDT            & 5.792  & 5.834  &       \\
                        CC3              & 5.923  & 5.935  & 6.017 \\
                        CASPT2           & 4.429  & 4.345  & 4.301 \\
                        CASPT2(IPEA)     & 4.583  & 4.509  & 4.472 \\
                        CASPT3           & 4.827  & 4.729  & 4.696 \\
                        CASPT3(IPEA)     & 4.792  & 4.703  & 4.672 \\
                        SC-NEVPT2        & 4.657  & 4.590  & 4.565 \\
                        PC-NEVPT2        & 4.665  & 4.593  & 4.566 \\
\hline
\mc{4}{c}{Beryllium, $^1D$, $\%T_1 = 31\%$} \\
                        TBE              & 8.038\Fp & 7.225\Fd & 7.151\Ft \\
                        FCI              & 8.038(0)& 7.225(0)& 7.151(0) \\
                        CCSDTQ           & 8.038  & 7.225  & 7.151 \\
                        CC4              & 8.035  & 7.225  & 7.151 \\
                        CCSDT            & 8.038  & 7.225  & 7.152 \\
                        CC3              & 8.041  & 7.234  & 7.158 \\
                        CASPT2           & 8.037  & 7.209  & 7.129 \\
                        CASPT2(IPEA)     & 8.036  & 7.212  & 7.137 \\
                        CASPT3           & 8.043  & 7.214  & 7.139 \\
                        CASPT3(IPEA)     & 8.042  & 7.215  & 7.142 \\
                        SC-NEVPT2        & 8.042  & 7.208  & 7.133 \\
                        PC-NEVPT2        & 8.042  & 7.208  & 7.133 \\
\hline
\mc{4}{c}{Borole, $^1A_1$, $\%T_1 = 81\%$} \\
                        TBE              & 6.579\Qp & 6.489\qdb & 6.484\Ttb \\
                        FCI              & 6.626(112) &    &       \\
                        CCSDTQ           & 6.579  &        &       \\
                        CC4              & 6.584  & 6.494  &       \\
                        CCSDT            & 6.637  & 6.546  & 6.541 \\
                        CC3              & 6.675  & 6.584  & 6.571 \\
                        CASPT2           & 6.201  & 6.022  & 5.979 \\
                        CASPT2(IPEA)     & 6.782  & 6.668  & 6.650 \\
                        CASPT3           & 6.738  & 6.630  & 6.622 \\
                        CASPT3(IPEA)     & 6.867  & 6.771  & 6.760 \\
                        SC-NEVPT2        & 6.628  & 6.506  & 6.484 \\
                        PC-NEVPT2        & 6.519  & 6.391  & 6.368 \\
\hline
\mc{4}{c}{Borole, $^1A_1$, $\%T_1 = 20\%$} \\
                        TBE              & 4.711\Qp & 4.702\qdb & 4.708\Ctb \\
			FCI              & \fk{4.722(49)} &     &       \\
                        CCSDTQ           & 4.711  &        &       \\
                        CC4              & 4.718  & 4.709  &       \\
                        CCSDT            & 5.069  & 5.098  & 5.211 \\
                        CC3              & 5.412  & 5.460  & 5.517 \\
                        CASPT2           & 4.628  & 4.566  & 4.580 \\
                        CASPT2(IPEA)     & 4.819  & 4.774  & 4.798 \\
                        CASPT3           & 4.765  & 4.747  & 4.753 \\
                        CASPT3(IPEA)     & 4.801  & 4.781  & 4.792 \\
                        SC-NEVPT2        & 4.800  & 4.758  & 4.785 \\
                        PC-NEVPT2        & 4.779  & 4.733  & 4.760 \\
\hline
\mc{4}{c}{Butadiene, $^1A_g$, $\%T_1 = 75\%$} \\
                        TBE              & 6.556\Qp & 6.506\qdb & 6.515\Ttb \\
                        FCI              & 6.560(47) & 6.530(46) & 6.633(186) \\
                        CCSDTQ           & 6.556  &        &       \\
                        CC4              & 6.558  & 6.508  &       \\
                        CCSDT            & 6.632  & 6.589  & 6.598 \\
                        CC3              & 6.731  & 6.678  & 6.671 \\
                        CASPT2           & 6.496  & 6.443  & 6.384 \\
                        CASPT2(IPEA)     & 6.791  & 6.777  & 6.736 \\
                        CASPT3           & 6.672  & 6.654  & 6.627 \\
                        CASPT3(IPEA)     & 6.753  & 6.748  & 6.718 \\
                        SC-NEVPT2        & 6.830  & 6.818  & 6.780 \\
                        PC-NEVPT2        & 6.754  & 6.738  & 6.700 \\
\hline
\mc{4}{c}{Carbon Dimer, $^1\Delta_g$, $\%T_1 = 0\%$} \\
                        TBE              & 2.292\Fp & 2.213\Fd & 2.091\Ft \\
                        FCI              & 2.292(0) & 2.213(0) & 2.091(0) \\
                        CCSDTQPH         & 2.292  &        &       \\
                        CCSDTQP          & 2.292  & 2.214  & 2.091 \\
                        CCSDTQ           & 2.316  & 2.240  & 2.127 \\
                        CC4              & 2.413  & 2.341  & 2.209 \\
                        CCSDT            & 2.694  & 2.632  & 2.567 \\
                        CC3              & 3.100  & 3.107  & 3.053 \\
                        CASPT2           & 2.457  & 2.418  & 2.277 \\
                        CASPT2(IPEA)     & 2.434  & 2.404  & 2.271 \\
                        CASPT3           & 2.308  & 2.242  & 2.096 \\
                        CASPT3(IPEA)     & 2.322  & 2.267  & 2.126 \\
                        SC-NEVPT2        & 2.347  & 2.281  & 2.144 \\
                        PC-NEVPT2        & 2.330  & 2.255  & 2.119 \\
\hline
\mc{4}{c}{Carbon Dimer, $^1\Sigma^+_g$, $\%T_1 = 0\%$} \\
                        TBE              & 2.509\Fp & 2.503\Fd & 2.420\Ft \\
                        FCI              & 2.509(1) & 2.503(1) & 2.420(0) \\
                        CCSDTQPH         & 2.510  &        &       \\
                        CCSDTQP          & 2.511  & 2.505  & 2.423 \\
                        CCSDTQ           & 2.524  & 2.521  & 2.447 \\
                        CC4              & 2.598  & 2.602  & 2.515 \\
                        CCSDT            & 2.847  & 2.874  & 2.861 \\
                        CC3              & 3.231  & 3.283  & 3.256 \\
                        CASPT2           & 2.627  & 2.650  & 2.518 \\
                        CASPT2(IPEA)     & 2.646  & 2.682  & 2.558 \\
                        CASPT3           & 2.515  & 2.513  & 2.397 \\
                        CASPT3(IPEA)     & 2.544  & 2.556  & 2.439 \\
                        SC-NEVPT2        & 2.585  & 2.596  & 2.478 \\
                        PC-NEVPT2        & 2.541  & 2.542  & 2.423 \\
\hline
\mc{4}{c}{Carbon Dimer, $^3\Sigma^-_g$, $\%T_1 = 0\%$} \\
                        TBE              & 1.394\Fp & 1.315\Fd & 1.258\Ft \\
                        FCI              & 1.394(1) & 1.315(1) & 1.258(2) \\
                        CCSDTQPH         & 1.394  &        &       \\
                        CCSDTQP          & 1.395  & 1.317  & 1.261 \\
                        CCSDTQ           & 1.431  & 1.355  & 1.308 \\
                        CCSDT            & 1.899  & 1.827  & 1.825 \\
                        CC3              & 2.394  & 2.399  & 2.388 \\
                        CASPT2           & 1.586  & 1.547  & 1.461 \\
                        CASPT2(IPEA)     & 1.587  & 1.559  & 1.476 \\
                        CASPT3           & 1.435  & 1.365  & 1.279 \\
                        CASPT3(IPEA)     & 1.459  & 1.402  & 1.318 \\
                        SC-NEVPT2        & 1.484  & 1.420  & 1.328 \\
                        PC-NEVPT2        & 1.432  & 1.357  & 1.266 \\
\hline
\mc{4}{c}{Carbon Trimer, $^1\Pi_g$, $\%T_1 = 77\%$} \\
                        TBE              & 4.042\Fp & 4.009\Fd & 4.008\Ft \\
                        FCI              & 4.042(3) & 4.009(3) & 4.008(4) \\
                        CCSDTQP          & 4.043  &        &       \\
                        CCSDTQ           & 4.051  & 4.017  & 4.014 \\
                        CC4              & 4.071  & 4.035  & 4.031 \\
                        CCSDT            & 4.128  & 4.093  & 4.095 \\
                        CC3              & 4.264  & 4.224  & 4.204 \\
                        CASPT2           & 3.779  & 3.699  & 3.654 \\
                        CASPT2(IPEA)     & 3.994  & 3.958  & 3.934 \\
                        CASPT3           & 4.011  & 3.968  & 3.969 \\
                        CASPT3(IPEA)     & 4.044  & 4.014  & 4.009 \\
                        SC-NEVPT2        & 4.087  & 4.077  & 4.063 \\
                        PC-NEVPT2        & 4.059  & 4.043  & 4.027 \\
\hline
\mc{4}{c}{Carbon Trimer, $^1\Delta_g$, $\%T_1 = 1\%$} \\
                        TBE              & 5.263\Fp & 5.222\Fd & 5.230\Ft \\
                        FCI              & 5.263(6) & 5.222(4) & 5.230(17) \\
                        CCSDTQP          & 5.270  &        &       \\
                        CCSDTQ           & 5.353  & 5.312  & 5.330 \\
                        CC4              & 5.412  & 5.376  & 5.373 \\
                        CCSDT            & 5.854  & 5.816  & 5.899 \\
                        CC3              & 6.653  & 6.646  & 6.681 \\
                        CASPT2           & 4.881  & 4.760  & 4.744 \\
                        CASPT2(IPEA)     & 5.081  & 5.015  & 5.027 \\
                        CASPT3           & 5.158  & 5.107  & 5.123 \\
                        CASPT3(IPEA)     & 5.198  & 5.156  & 5.172 \\
                        SC-NEVPT2        & 5.213  & 5.191  & 5.209 \\
                        PC-NEVPT2        & 5.255  & 5.236  & 5.255 \\
\hline
\mc{4}{c}{Carbon Trimer, $^1\Sigma^+_g$, $\%T_1 = 1\%$} \\
                        TBE              & 5.924\Fp & 5.897\Fd & 5.908\Ft \\
                        FCI              & 5.924(6) & 5.897(5) & 5.908(11) \\
                        CCSDTQP          & 5.932  &        &       \\
                        CCSDTQ           & 6.021  & 5.996  & 6.016 \\
                        CC4              & 6.096  & 6.073  & 6.071 \\
                        CCSDT            & 6.520  & 6.493  & 6.571 \\
                        CC3              & 7.196  & 7.194  & 7.238 \\
                        CASPT2           & 5.602  & 5.521  & 5.506 \\
                        CASPT2(IPEA)     & 5.817  & 5.769  & 5.776 \\
                        CASPT3           & 5.860  & 5.813  & 5.826 \\
                        CASPT3(IPEA)     & 5.919  & 5.884  & 5.896 \\
                        SC-NEVPT2        & 5.977  & 5.974  & 5.993 \\
                        PC-NEVPT2        & 5.966  & 5.969  & 5.986 \\
\hline
\mc{4}{c}{Criegee's Intermediate, $^1A''$, $\%T_1 = 80\%$} \\
                        TBE              & 2.474\Fp & 2.443\Fd & 2.403\qtb \\
                        FCI              & 2.474(6) & 2.443(3) & 2.377(57) \\
                        CCSDTQ           & 2.469  & 2.442  &       \\
                        CC4              & 2.471  & 2.445  & 2.405 \\
                        CCSDT            & 2.471  & 2.446  & 2.408 \\
                        CC3              & 2.460  & 2.426  & 2.384 \\
                        CASPT2           & 2.318  & 2.291  & 2.234 \\
                        CASPT2(IPEA)     & 2.391  & 2.371  & 2.318 \\
                        CASPT3           & 2.424  & 2.391  & 2.339 \\
                        CASPT3(IPEA)     & 2.428  & 2.400  & 2.348 \\
                        SC-NEVPT2        & 2.449  & 2.440  & 2.388 \\
                        PC-NEVPT2        & 2.501  & 2.488  & 2.440 \\
\hline
\mc{4}{c}{Criegee's Intermediate, $^1A'$, $\%T_1 = 80\%$} \\
                        TBE              & 3.807\Fp & 3.747\Fd & 3.715\qtb \\
                        FCI              & 3.807(8) & 3.747(12) & 3.721(54) \\
                        CCSDTQ           & 3.808  & 3.742  &       \\
                        CC4              & 3.804  & 3.738  & 3.706 \\
                        CCSDT            & 3.829  & 3.765  & 3.737 \\
                        CC3              & 3.831  & 3.760  & 3.725 \\
                        CASPT2           & 3.766  & 3.703  & 3.645 \\
                        CASPT2(IPEA)     & 3.977  & 3.932  & 3.881 \\
                        CASPT3           & 3.918  & 3.863  & 3.825 \\
                        CASPT3(IPEA)     & 3.963  & 3.914  & 3.870 \\
                        SC-NEVPT2        & 3.971  & 3.923  & 3.866 \\
                        PC-NEVPT2        & 3.856  & 3.801  & 3.746 \\
\hline
\mc{4}{c}{Cyclobutadiene, $^1A_g$, $\%T_1 = 1\%$} \\
                        TBE              & 4.073\Qp & 4.046\qdb & 4.036\Ctb \\
                        FCI              & 4.084(12) & 4.04(3) & 4.03(9) \\
                        CCSDTQ           & 4.073  &        &       \\
                        CC4              & 4.067  & 4.040  &       \\
                        CCSDT            & 4.311  & 4.327  & 4.429 \\
                        CC3              & 4.658  & 4.711  & 4.777 \\
                        CASPT2           & 4.040  & 3.982  & 3.971 \\
                        CASPT2(IPEA)     & 4.096  & 4.047  & 4.039 \\
                        CASPT3           & 4.077  & 4.041  & 4.031 \\
                        CASPT3(IPEA)     & 4.087  & 4.050  & 4.041 \\
                        SC-NEVPT2        & 4.130  & 4.093  & 4.086 \\
                        PC-NEVPT2        & 4.103  & 4.064  & 4.056 \\
\hline
\mc{4}{c}{Cyclopentadiene, $^1A_1$, $\%T_1 = 79\%$} \\
                        TBE              & 6.522\Qp & 6.459\qdb & 6.451\Ttb \\
                        FCI              & 6.572(46) &     &       \\
                        CCSDTQ           & 6.522  &        &       \\
                        CC4              & 6.526  & 6.463  &       \\
                        CCSDT            & 6.597  & 6.531  & 6.523 \\
                        CC3              & 6.672  & 6.595  & 6.570 \\
                        CASPT2           & 6.383  & 6.303  & 6.249 \\
                        CASPT2(IPEA)     & 6.755  & 6.721  & 6.686 \\
                        CASPT3           & 6.619  & 6.591  & 6.558 \\
                        CASPT3(IPEA)     & 6.706  & 6.688  & 6.650 \\
                        SC-NEVPT2        & 6.899  & 6.878  & 6.848 \\
                        PC-NEVPT2        & 6.872  & 6.848  & 6.817 \\
\hline
\mc{4}{c}{Cyclopentadienethione, $^1A_1$, $\%T_1 = 52\%$} \\
                        TBE              & 5.366\qp & 5.363\Cdb & 5.329\Ctb \\
                        FCI              & 5.584(490) &    &       \\
                        CC4              & 5.366  &        &       \\
                        CCSDT            & 5.678  & 5.711  & 5.744 \\
                        CC3              & 5.860  & 5.895  & 5.884 \\
                        CASPT2           & 5.042  & 4.983  & 4.919 \\
                        CASPT2(IPEA)     & 5.461  & 5.458  & 5.418 \\
                        CASPT3           & 5.377  & 5.373  & 5.340 \\
                        CASPT3(IPEA)     & 5.452  & 5.460  & 5.423 \\
                        SC-NEVPT2        & 5.496  & 5.498  & 5.466 \\
                        PC-NEVPT2        & 5.462  & 5.462  & 5.428 \\
\hline
\mc{4}{c}{Cyclopentadienethione, $^1B_1$, $\%T_1 = 1\%$} \\
                        TBE              & 3.211\Fp & 3.195\Cdb & 3.156\Ctb \\
                        FCI              & 3.211(21) &     &       \\
                        CC4              & 3.300  &        &       \\
                        CCSDT            & 3.842  & 3.929  & 4.036 \\
                        CC3              & 4.266  & 4.349  & 4.399 \\
                        CASPT2           & 3.020  & 2.966  & 2.900 \\
                        CASPT2(IPEA)     & 3.197  & 3.185  & 3.137 \\
                        CASPT3           & 3.182  & 3.166  & 3.127 \\
                        CASPT3(IPEA)     & 3.222  & 3.216  & 3.178 \\
                        SC-NEVPT2        & 3.222  & 3.214  & 3.169 \\
                        PC-NEVPT2        & 3.211  & 3.200  & 3.154 \\
\hline
\mc{4}{c}{Cyclopentadienethione, $^1A_1$, $\%T_1 = 3\%$} \\
                        TBE              & 5.780\qp & 5.615\Ndb & 5.555\Ntb \\
                        FCI              & 5.901(213) &    &       \\
                        CC4              & 5.780  &        &       \\
                        CCSDT            & 6.376  & 6.318  & 6.365 \\
                        CC3              & 6.694  & 6.627  & 6.622 \\
                        CASPT2           & 5.414  & 5.156  & 4.995 \\
                        CASPT2(IPEA)     & 5.879  & 5.747  & 5.679 \\
                        CASPT3           & 6.110  & 5.966  & 5.942 \\
                        CASPT3(IPEA)     & 6.182  & 6.057  & 6.032 \\
                        SC-NEVPT2        & 5.601  & 5.451  & 5.397 \\
                        PC-NEVPT2        & 5.513  & 5.348  & 5.288 \\
\hline
\mc{4}{c}{Cyclopentadienethione, $^3B_1$, $\%T_1 = 4\%$} \\
                        TBE              & 3.170\Fp & 3.152\Cdb & 3.115\Ctb \\
                        FCI              & 3.170(26) &     &       \\
                        CCSDT            & 3.815  & 3.900  &       \\
                        CC3              & 4.254  & 4.349  & 4.385 \\
                        CASPT2           & 2.984  & 2.929  & 2.867 \\
                        CASPT2(IPEA)     & 3.163  & 3.149  & 3.105 \\
                        CASPT3           & 3.141  & 3.124  & 3.087 \\
                        CASPT3(IPEA)     & 3.183  & 3.176  & 3.140 \\
                        SC-NEVPT2        & 3.189  & 3.179  & 3.136 \\
                        PC-NEVPT2        & 3.178  & 3.165  & 3.121 \\
\hline
\mc{4}{c}{Cyclopentadienone, $^1A_1$, $\%T_1 = 50\%$} \\
                        TBE              & 6.850\qp & 6.798\Ndb & 6.714\Ntb \\
                        FCI              & 7.005(183) &    &       \\
                        CC4              & 6.850  &        &       \\
                        CCSDT            & 7.066  & 6.957  & 6.967 \\
                        CC3              & 7.202  & 7.118  & 7.098 \\
                        CASPT2           & 6.477  & 6.350  & 6.244 \\
                        CASPT2(IPEA)     & 7.323  & 7.253  & 7.172 \\
                        CASPT3           & 7.557  & 7.479  & 7.468 \\
                        CASPT3(IPEA)     & 7.642  & 7.572  & 7.548 \\
                        SC-NEVPT2        & 7.004  & 6.953  & 6.872 \\
                        PC-NEVPT2        & 6.797  & 6.746  & 6.662 \\
\hline
\mc{4}{c}{Cyclopentadienone, $^1B_1$, $\%T_1 = 3\%$} \\
                        TBE              & 5.103\Fp & 5.038\Cdb & 5.009\Ctb \\
                        FCI              & 5.103(22) &     &       \\
                        CC4              & 5.098  &        &       \\
                        CCSDT            & 5.665  & 5.689  & 5.824 \\
                        CC3              & 6.057  & 6.072  & 6.125 \\
                        CASPT2           & 4.777  & 4.691  & 4.649 \\
                        CASPT2(IPEA)     & 5.006  & 4.944  & 4.914 \\
                        CASPT3           & 5.102  & 5.037  & 5.008 \\
                        CASPT3(IPEA)     & 5.107  & 5.047  & 5.018 \\
                        SC-NEVPT2        & 5.111  & 5.062  & 5.041 \\
                        PC-NEVPT2        & 5.102  & 5.046  & 5.025 \\
\hline
\mc{4}{c}{Cyclopentadienone, $^1A_1$, $\%T_1 = 74\%$} \\
                        TBE              & 5.826\qp & 5.819\Cdb & 5.795\Ctb \\
                        FCI              & 5.750(348) &    &       \\
                        CC4              & 5.826  &        &       \\
                        CCSDT            & 6.119  & 6.113  & 6.126 \\
                        CC3              & 6.257  & 6.231  & 6.205 \\
                        CASPT2           & 5.720  & 5.664  & 5.612 \\
                        CASPT2(IPEA)     & 6.044  & 6.019  & 5.983 \\
                        CASPT3           & 5.940  & 5.933  & 5.910 \\
                        CASPT3(IPEA)     & 6.011  & 6.007  & 5.979 \\
                        SC-NEVPT2        & 6.103  & 6.088  & 6.057 \\
                        PC-NEVPT2        & 6.071  & 6.053  & 6.022 \\
\hline
\mc{4}{c}{Cyclopentadienone, $^3B_1$, $\%T_1 = 10\%$} \\
                        TBE              & 4.910\Fp & 4.844\Cdb & 4.821\Ctb \\
                        FCI              & 4.910(39) &     &       \\
                        CCSDT            & 5.553  & 5.574  &       \\
                        CC3              & 5.979  & 5.990  & 6.046 \\
                        CASPT2           & 4.669  & 4.579  & 4.542 \\
                        CASPT2(IPEA)     & 4.905  & 4.839  & 4.814 \\
                        CASPT3           & 4.986  & 4.920  & 4.897 \\
                        CASPT3(IPEA)     & 4.997  & 4.936  & 4.913 \\
                        SC-NEVPT2        & 5.002  & 4.948  & 4.933 \\
                        PC-NEVPT2        & 4.988  & 4.929  & 4.913 \\
\hline
\mc{4}{c}{Diazete, $^1A_1$, $\%T_1 = 1\%$} \\
                        TBE              & 6.726\Qp & 6.635\qdb & 6.605\Ctb \\
                        FCI              & 6.727(47) & 6.651(76) & 6.655(115) \\
                        CCSDTQ           & 6.726  &        &       \\
                        CC4              & 6.722  & 6.631  &       \\
                        CCSDT            & 7.097  & 7.080  & 7.192 \\
                        CC3              & 7.465  & 7.455  & 7.509 \\
                        CASPT2           & 6.729  & 6.623  & 6.584 \\
                        CASPT2(IPEA)     & 6.824  & 6.731  & 6.699 \\
                        CASPT3           & 6.787  & 6.702  & 6.672 \\
                        CASPT3(IPEA)     & 6.796  & 6.712  & 6.684 \\
                        SC-NEVPT2        & 6.846  & 6.765  & 6.732 \\
                        PC-NEVPT2        & 6.776  & 6.689  & 6.649 \\
\hline
\mc{4}{c}{Ethylene, $^1A_g$, $\%T_1 = 20\%$} \\
                        TBE              & 13.387\Fp & 13.068\Fd & 12.899\Qtb \\
                        FCI              & 13.387(3) & 13.068(10) & 12.898(29) \\
                        CCSDTQP          & 13.386 &        &       \\
                        CCSDTQ           & 13.390 & 13.068 & 12.899\\
                        CC4              & 13.391 & 13.076 & 12.906\\
                        CCSDT            & 13.497 & 13.204 & 13.080\\
                        CC3              & 13.820 & 13.569 & 13.420\\
                        CASPT2           & 13.418 & 13.163 & 13.093\\
                        CASPT2(IPEA)     & 13.474 & 13.230 & 13.163\\
                        CASPT3           & 13.475 & 13.230 & 13.177\\
                        CASPT3(IPEA)     & 13.499 & 13.257 & 13.205\\
                        SC-NEVPT2        & 13.570 & 13.334 & 13.262\\
                        PC-NEVPT2        & 14.353 & 13.418 & 13.109\\
\hline
\mc{4}{c}{Formaldehyde, $^1A_1$, $\%T_1 = 5\%$} \\
                        TBE              & 10.859\Fp & 10.422\Fd & 10.426\Ft \\
                        FCI              & 10.859(1) & 10.422(34) & 10.426(12) \\
                        CCSDTQP          & 10.860 &        &       \\
                        CCSDTQ           & 10.872 & 10.435 & 10.428\\
                        CC4              & 10.884 & 10.552 & 10.434\\
                        CCSDT            & 11.102 & 10.784 & 10.785\\
                        CC3              & 11.492 & 11.219 & 11.200\\
                        CASPT2           & 10.923 & 10.500 & 10.400\\
                        CASPT2(IPEA)     & 10.948 & 10.522 & 10.420\\
                        CASPT3           & 11.182 & 10.788 & 10.718\\
                        CASPT3(IPEA)     & 11.175 & 10.782 & 10.713\\
                        SC-NEVPT2        & 10.865 & 10.398 & 10.296\\
                        PC-NEVPT2        & 10.840 & 10.370 & 10.265\\
\hline
\mc{4}{c}{Glyoxal, $^1A_g$, $\%T_1 = 1\%$} \\
			TBE              & \fk{5.628}\Fp & \fk{5.522}\qdb & \fk{5.492}\Ctb \\
			FCI              & \fk{5.628(34)} & 5.513(75) & \\
                        CCSDTQ           & 5.670  &        &       \\
                        CC4              & 5.699  & 5.593  &       \\
                        CCSDT            & 6.243  & 6.222  & 6.353 \\
                        CC3              & 6.735  & 6.706  & 6.763 \\
                        CASPT2           & 5.406  & 5.270  & 5.211 \\
                        CASPT2(IPEA)     & 5.544  & 5.421  & 5.372 \\
                        CASPT3           & 5.689  & 5.580  & 5.549 \\
                        CASPT3(IPEA)     & 5.672  & 5.564  & 5.535 \\
                        SC-NEVPT2        & 5.678  & 5.581  & 5.546 \\
                        PC-NEVPT2        & 5.660  & 5.557  & 5.518 \\
\hline
\mc{4}{c}{Hexatriene, $^1A_g$, $\%T_1 = 65\%$} \\
                        TBE              & 5.471\qp & 5.457\Cdb & 5.435\Ctb \\
                        CC4              & 5.471  &        &       \\
                        CCSDT            & 5.617  & 5.627  & 5.652 \\
                        CC3              & 5.759  & 5.755  & 5.747 \\
                        CASPT2           & 5.277  & 5.222  & 5.171 \\
                        CASPT2(IPEA)     & 5.617  & 5.603  & 5.571 \\
                        CASPT3           & 5.508  & 5.494  & 5.472 \\
                        CASPT3(IPEA)     & 5.586  & 5.582  & 5.556 \\
                        SC-NEVPT2        & 5.702  & 5.695  & 5.667 \\
                        PC-NEVPT2        & 5.673  & 5.664  & 5.636 \\
\hline
\mc{4}{c}{Naphthalene, $^1A_g$, $\%T_1 = 72\%$} \\
                        TBE              & 6.814\Cp & 6.767\Cd & 6.748\Ct \\
                        CCSDT            & 6.993  & 6.935  &       \\
                        CC3              & 6.987  & 6.909  & 6.868 \\
                        CASPT2           & 6.201  & 6.101  & 5.942 \\
                        CASPT2(IPEA)     & 6.888  & 6.852  & 6.793 \\
                        CASPT3           & 6.814  & 6.767  & 6.748 \\
                        CASPT3(IPEA)     & 6.887  & 6.858  & 6.809 \\
                        SC-NEVPT2        & 7.038  & 7.011  & 6.958 \\
                        PC-NEVPT2        & 6.985  & 6.953  & 6.900 \\
\hline
\mc{4}{c}{Nitrosomethane, $^1A'$, $\%T_1 = 3\%$} \\
                        TBE              & 4.861\Fp & 4.816\Fp & 4.732\qtb \\
                        FCI              & 4.861(2) & 4.816(12) & 4.764(34) \\
                        CCSDTQ           & 4.895  & 4.848  &       \\
                        CC4              & 4.926  & 4.878  & 4.794 \\
                        CCSDT            & 5.260  & 5.258  & 5.293 \\
                        CC3              & 5.729  & 5.749  & 5.757 \\
                        CASPT2           & 4.908  & 4.862  & 4.775 \\
                        CASPT2(IPEA)     & 4.918  & 4.873  & 4.787 \\
                        CASPT3           & 4.859  & 4.820  & 4.742 \\
                        CASPT3(IPEA)     & 4.866  & 4.825  & 4.748 \\
                        SC-NEVPT2        & 4.938  & 4.901  & 4.815 \\
                        PC-NEVPT2        & 4.923  & 4.881  & 4.794 \\
\hline
\mc{4}{c}{Nitrous Acid, $^1A'$, $\%T_1 = 50\%$} \\
			TBE              & \fk{8.170\Fd} & \fk{8.057\Fd} & \fk{7.969\Ft} \\
			FCI              & 8.170(40) & \fk{8.057(32)} & \fk{7.969(36)} \\
                        CCSDTQ           & 8.185  & 8.081  &       \\
                        CC4              & 8.198  & 8.104  & 7.993 \\
                        CCSDT            & 8.500  & 8.500  & 8.524 \\
                        CC3              & 8.915  & 9.174  & 9.105 \\
                        CASPT2           & 8.097  & 8.089  & 7.970 \\
                        CASPT2(IPEA)     & 8.089  & 8.084  & 7.967 \\
                        CASPT3           & 8.064  & 8.052  & 7.945 \\
                        CASPT3(IPEA)     & 8.059  & 8.047  & 7.939 \\
                        SC-NEVPT2        & 8.106  & 8.109  & 7.992 \\
                        PC-NEVPT2        & 8.126  & 8.121  & 8.001 \\
\hline
\mc{4}{c}{Nitroxyl, $^1A'$, $\%T_1 = 0\%$} \\
                        TBE              & 4.511\Fp & 4.397\Fp & 4.333\Fp \\
                        FCI              & 4.511(1) & 4.397(1) & 4.333(1) \\
                        CCSDTQP          & 4.513  & 4.399  &       \\
                        CCSDTQ           & 4.535  & 4.424  & 4.364 \\
                        CC4              & 4.562  & 4.454  & 4.380 \\
                        CCSDT            & 4.819  & 4.756  & 4.785 \\
                        CC3              & 5.278  & 5.247  & 5.257 \\
                        CASPT2           & 4.549  & 4.447  & 4.358 \\
                        CASPT2(IPEA)     & 4.554  & 4.454  & 4.366 \\
                        CASPT3           & 4.527  & 4.428  & 4.361 \\
                        CASPT3(IPEA)     & 4.530  & 4.432  & 4.364 \\
                        SC-NEVPT2        & 4.585  & 4.485  & 4.398 \\
                        PC-NEVPT2        & 4.563  & 4.457  & 4.371 \\
\hline
\mc{4}{c}{Octatetraene, $^1A_g$, $\%T_1 = 64\%$} \\
                        TBE              & 4.709\Cp & 4.694\Cd & 4.680\Ct \\
                        CCSDT            & 4.907  & 4.929  &       \\
                        CC3              & 5.048  & 5.048  & 5.042 \\
                        CASPT2           & 4.424  & 4.375  & 4.326 \\
                        CASPT2(IPEA)     & 4.780  & 4.771  & 4.741 \\
                        CASPT3           & 4.709  & 4.694  & 4.680 \\
                        CASPT3(IPEA)     & 4.775  & 4.772  & 4.751 \\
                        SC-NEVPT2        & 4.843  & 4.840  & 4.814 \\
                        PC-NEVPT2        & 4.816  & 4.810  & 4.783 \\
\hline
\mc{4}{c}{Oxalyl Fluoride, $^1A_1$, $\%T_1 = 2\%$} \\
                        TBE              &  9.056\Cp &  8.976\Cd &  8.923\Ct\\
                        CC4              &  9.206 &  9.126 &       \\
                        CCSDT            &  9.861 &  9.874 & 10.028\\
                        CC3              & 10.257 & 10.252 & 10.305\\
                        CASPT2           &  8.813 &  8.718 &  8.643\\
                        CASPT2(IPEA)     &  8.963 &  8.878 &  8.814\\
                        CASPT3           &  9.056 &  8.976 &  8.923\\
                        CASPT3(IPEA)     &  9.029 &  8.949 &  8.899\\
                        SC-NEVPT2        &  9.075 &  9.009 &  8.958\\
                        PC-NEVPT2        &  9.070 &  8.997 &  8.942\\
\hline
\mc{4}{c}{Pyrazine, $^1A_g$, $\%T_1 = 71\%$} \\
                        TBE              & 8.582\qp & 8.560\qd & 8.480\Ctb \\
                        FCI              & 8.656(107) &    &       \\
                        CC4              & 8.582  & 8.560  &       \\
                        CCSDT            & 8.858  & 8.775  & 8.693 \\
                        CC3              & 8.879  & 8.771  & 8.697 \\
                        CASPT2           & 8.419  & 8.311  & 8.208 \\
                        CASPT2(IPEA)     & 8.884  & 8.824  & 8.746 \\
                        CASPT3           & 8.754  & 8.704  & 8.624 \\
                        CASPT3(IPEA)     & 8.800  & 8.759  & 8.676 \\
                        SC-NEVPT2        & 9.163  & 9.118  & 9.051 \\
                        PC-NEVPT2        & 9.119  & 9.066  & 8.999 \\
\hline
\mc{4}{c}{Pyrazine, $^1A_g$, $\%T_1 = 12\%$} \\
                        TBE              & 8.049\qp & 7.986\qd & 7.904\Ctb \\
                        CC4              & 8.049  & 7.986  &       \\
                        CCSDT            & 8.794  & 8.697  & 8.813 \\
                        CC3              & 9.274  & 9.172  & 9.168 \\
                        CASPT2           & 7.842  & 7.687  & 7.584 \\
                        CASPT2(IPEA)     & 8.074  & 7.938  & 7.845 \\
                        CASPT3           & 8.147  & 8.031  & 7.949 \\
                        CASPT3(IPEA)     & 8.145  & 8.028  & 7.945 \\
                        SC-NEVPT2        & 8.275  & 8.151  & 8.067 \\
                        PC-NEVPT2        & 8.251  & 8.120  & 8.037 \\
\hline
\mc{4}{c}{Tetrazine, $^1A_g$, $\%T_1 = 1\%$} \\
                        TBE              & 5.035\Cp & 4.991\Cd & 4.951\Ct \\
			FCI              & \fk{4.991(73)} &     &       \\
                        CC4              & 5.063  & 4.970  &       \\
                        CCSDT            & 5.856  & 5.858  & 5.958 \\
                        CC3              & 6.222  & 6.218  & 6.209 \\
                        CASPT2           & 4.497  & 4.428  & 4.318 \\
                        CASPT2(IPEA)     & 4.724  & 4.666  & 4.566 \\
                        CASPT3           & 5.035  & 4.991  & 4.951 \\
                        CASPT3(IPEA)     & 5.009  & 4.963  & 4.917 \\
                        SC-NEVPT2        & 4.823  & 4.778  & 4.688 \\
                        PC-NEVPT2        & 4.749  & 4.699  & 4.608 \\
\hline
\mc{4}{c}{Tetrazine, $^1B_{3g}$, $\%T_1 = 1\%$} \\
                        TBE              & 6.282\Cp & 6.250\Cd & 6.215\Ct \\
			FCI              & \fk{6.191(111)} &    &       \\
                        CC4              & 6.308  & 6.246  &       \\
                        CCSDT            & 7.300  & 7.302  & 7.434 \\
                        CC3              & 7.640  & 7.618  & 7.617 \\
                        CASPT2           & 5.437  & 5.340  & 5.218 \\
                        CASPT2(IPEA)     & 5.997  & 5.945  & 5.853 \\
                        CASPT3           & 6.282  & 6.250  & 6.215 \\
                        CASPT3(IPEA)     & 6.279  & 6.247  & 6.201 \\
                        SC-NEVPT2        & 6.299  & 6.273  & 6.200 \\
                        PC-NEVPT2        & 6.255  & 6.223  & 6.149 \\
\hline
\mc{4}{c}{Tetrazine, $^3B_{3g}$, $\%T_1 = 6\%$} \\
                        TBE              & 5.919\Cp & 5.866\Cd & 5.848\Ct \\
			FCI              & \fk{5.894(103)} &    &       \\
                        CCSDT            & 6.944  & 6.931  &       \\
                        CC3              & 7.355  & 7.331  & 7.347 \\
                        CASPT2           & 5.079  & 4.968  & 4.861 \\
                        CASPT2(IPEA)     & 5.542  & 5.471  & 5.388 \\
                        CASPT3           & 5.919  & 5.866  & 5.848 \\
                        CASPT3(IPEA)     & 5.909  & 5.857  & 5.829 \\
                        SC-NEVPT2        & 5.688  & 5.642  & 5.570 \\
                        PC-NEVPT2        & 5.632  & 5.579  & 5.506 \\
\end{longtable}
\end{squeezetable}

\begin{squeezetable}
\begin{table*}
\caption{Present TBEs (in \si{\eV}), the composite method used to compute them, the symmetry and doubly-excited character of the state (P stands for partial and G for genuine),
whether they are considered to be safe (uncertainty less than $\SI{0.043}{\eV}$), and their comparison to previous TBEs,
ordered by the number of valence electrons.
States with blank entries in the previous and difference columns represent new additions to the \textsc{quest} database.
For the sake of conciseness, the 6-31+G*, aug-cc-pVDZ, and aug-cc-pVTZ basis sets are labeled P, D, and T, respectively.}
\label{tab:tbe}
\begin{ruledtabular}
\begin{tabular}{dllccdddl}
\mc{1}{c}{No. $e^-$} & Molecule & State & P/G & Safe & \mc{1}{c}{New} & \mc{1}{c}{Prev.} & \mc{1}{c}{Diff.} & Method \\ 
\hline
 4 & Beryllium              & $^1D$          & P & Y &  7.151 &  7.15  \cite{Veril_2021}   & +0.001  & exFCI/T \\ 
 8 & Carbon dimer           & $^1\Delta_g$   & G & Y &  2.091 &  2.09  \cite{Veril_2021}   & +0.001  & exFCI/T \\ 
   &                        & $^1\Sigma^+_g$ & G & Y &  2.420 &  2.42  \cite{Veril_2021}   &  0.000  & exFCI/T \\ 
   &                        & $^3\Sigma^-_g$ & G & Y &  1.258 &                            &         & exFCI/T \\
12 & Nitroxyl               & $^1A'$         & G & Y &  4.333 &  4.33  \cite{Veril_2021}   & +0.003  & exFCI/T \\ 
12 & Formaldehyde           & $^1A_1$        & G & Y & 10.426 & 10.35  \cite{Veril_2021}   & +0.076  & exFCI/T \\ 
12 & Ethylene               & $^1A_g$        & G & Y & 12.899 & 12.92  \cite{Veril_2021}   & -0.021  & exFCI/T \\ 
12 & Carbon trimer          & $^1\Pi_g$      & P & Y &  4.008 &                            &         & exFCI/T \\
   &                        & $^1\Delta_g$   & G & Y &  5.230 &  5.22  \cite{Veril_2021}   & +0.010  & exFCI/T \\ 
   &                        & $^1\Sigma^+_g$ & G & Y &  5.908 &  5.91  \cite{Veril_2021}   & -0.002  & exFCI/T \\ 
18 & Criegee's intermediate & $^1A''$        & P & Y &  2.403 &                            &         & exFCI/D + CC4/T $-$ CC4/D \\
   &                        & $^1A'$         & P & Y &  3.715 &                            &         & exFCI/D + CC4/T $-$ CC4/D \\
18 & Nitrosomethane         & $^1A'$         & G & Y &  4.732 &  4.76  \cite{Veril_2021}   & -0.028  & exFCI/D + CASPT3/T $-$ CASPT3/D \\ 
18 & Nitrous acid           & $^1A_1$        & G & Y &  7.969 &                            &         & exFCI/T \\
20 & Diazete                & $^1A_1$        & G & Y &  6.605 &                            &         & CCSDTQ/P + CC4/D $-$ CC4/P + CASPT3/D $-$ CASPT3/P \\
20 & Cyclobutadiene         & $^1A_g$        & G & Y &  4.036 &  4.038 \cite{Monino_2022a} & -0.002  & CCSDTQ/P + CC4/D $-$ CC4/P + CASPT3/T $-$ CASPT3/D \\ 
22 & Butadiene              & $^1A_g$        & P & Y &  6.515 &  6.50  \cite{Veril_2021}   & +0.015  & CCSDTQ/P + CC4/D $-$ CC4/P + CCSDT/T $-$ CCSDT/D \\ 
22 & Acrolein               & $^1A'$         & P & Y &  7.928 &  7.929 \cite{Loos_2022b}   & -0.001  & CCSDTQ/P + CC4/D $-$ CC4/P + CCSDT/T $-$ CCSDT/D \\ 
22 & Glyoxal                & $^1A_g$        & G & N &  5.492 &  5.61  \cite{Veril_2021}   & -0.118  & exFCI/P + CC4/D $-$ CC4/P + CASPT3/T $-$ CASPT3/D \\ 
24 & Borole                 & $^1A_1$        & G & N &  4.708 &                            &         & exFCI/P + CASPT3/T $-$ CASPT3/P \\
   &                        & $^1A_1$        & P & Y &  6.484 &                            &         & CCSDTQ/P + CC4/D $-$ CC4/P + CCSDT/T $-$ CCSDT/D \\
26 & Cyclopentadiene        & $^1A_1$        & P & Y &  6.451 &  6.452 \cite{Loos_2022b}   & -0.001  & CCSDTQ/P + CC4/D $-$ CC4/P + CCSDT/T $-$ CCSDT/P \\ 
30 & Benzene                & $^1E_{2g}$     & P & N &  8.190 &  8.28  \cite{Veril_2021}   & -0.090  & CC4/D + CCSDT/T $-$ CCSDT/D \\ 
   &                        & $^1A_{1g}$     & G & N & 10.315 & 10.55  \cite{Veril_2021}   & -0.235  & CASPT3/T \\ 
30 & Cyclopentadienethione  & $^1B_1$        & G & N &  3.156 &  3.154 \cite{Veril_2021}   & +0.002  & exFCI/P + CASPT3/T $-$ CASPT3/P \\ 
   &                        & $^1A_1$        & P & N &  5.329 &  5.428 \cite{Veril_2021}   & +0.127  & CC4/P + CASPT3/T $-$ CASPT3/P \\ 
   &                        & $^1A_1$        & G & N &  5.555 &                            &         & CC4/P + PC-NEVPT2/T $-$ PC-NEVPT2/P \\
   &                        & $^3B_1$        & G & N &  3.115 &  3.121 \cite{Veril_2021}   & -0.006  & exFCI/P + CASPT3/T $-$ CASPT3/P \\ 
30 & Cyclopentadienone      & $^1B_1$        & G & N &  5.009 &  5.025 \cite{Veril_2021}   & -0.016  & exFCI/P + CASPT3/T $-$ CASPT3/P \\ 
   &                        & $^1A_1$        & G & N &  5.795 &  6.087 \cite{Veril_2021}   & -0.292  & CC4/P + CASPT3/T $-$ CASPT3/P \\ 
   &                        & $^1A_1$        & P & N &  6.714 &  6.662 \cite{Veril_2021}   & -0.052  & CC4/P + PC-NEVPT2/T $-$ PC-NEVPT2/P \\ 
   &                        & $^3B_1$        & G & N &  4.821 &  4.913 \cite{Veril_2021}   & -0.092  & exFCI/P + CASPT3/T $-$ CASPT3/P \\ 
30 & Pyrazine               & $^1A_g$        & G & N &  7.904 &  8.037 \cite{Veril_2021}   & -0.133  & CC4/D + CASPT3/T $-$ CASPT3/D \\ 
   &                        & $^1A_g$        & P & N &  8.480 &  8.697 \cite{Veril_2021}   & -0.217  & CC4/D + CASPT3/T $-$ CASPT3/D \\ 
30 & Tetrazine              & $^1A_g$        & G & N &  4.951 &  4.608 \cite{Veril_2021}   & +0.343  & CASPT3/T \\ 
   &                        & $^1B_{3g}$     & G & N &  6.215 &  6.149 \cite{Veril_2021}   & +0.066  & CASPT3/T \\ 
   &                        & $^3B_{3g}$     & G & N &  5.848 &  5.506 \cite{Veril_2021}   & +0.342  & CASPT3/T \\ 
32 & Hexatriene             & $^1A_g$        & P & N &  5.435 &  5.459 \cite{Loos_2022b}   & -0.024  & CC4/P + CASPT3/T $-$ CASPT3/P \\ 
34 & Oxalyl fluoride        & $^1A_g$        & G & N &  8.923 &                            &         & CASPT3/T \\
40 & Benzoquinone           & $^1A_g$        & G & N &  4.566 &  4.566 \cite{Veril_2021}   &  0.000  & PC-NEVPT2/T \\ 
   &                        & $^1A_g$        & P & N &  6.351 &                            &         & CC4/P + CCSDT/D $-$ CCSDT/P + CC3/T $-$ CC3/D \\
   &                        & $^1B_{3u}$     & P & Y &  5.656 &  5.796 \cite{Veril_2021}   & -0.140  & CC4/P + CCSDT/D $-$ CCSDT/P + CC3/T $-$ CC3/D \\ 
   &                        & $^1B_{2g}$     & P & Y &  5.764 &  5.970 \cite{Veril_2021}   & -0.206  & CC4/P + CCSDT/D $-$ CCSDT/P + CC3/T $-$ CC3/D \\ 
   &                        & $^1A_u$        & P & Y &  6.083 &  6.346 \cite{Veril_2021}   & -0.263  & CC4/P + CCSDT/D $-$ CCSDT/P + CC3/T $-$ CC3/D \\ 
   &                        & $^1B_{1g}$     & P & N &  6.469 &                            &         & CC4/P + CCSDT/D $-$ CCSDT/P + CC3/T $-$ CC3/D \\
42 & Octatetraene           & $^1A_g$        & P & N &  4.680 &  4.901 \cite{Veril_2021}   & -0.221  & CASPT3/T \\ 
48 & Naphthalene            & $^1A_g$        & P & N &  6.748 &  6.874 \cite{Veril_2021}   & -0.126  & CASPT3/T \\ 
\end{tabular}
\end{ruledtabular}
\end{table*}
\end{squeezetable}

\subsection{Beryllium, Carbon Dimer, and Carbon Trimer}
\label{sec:ii}

Because it is a very small, four-electron system, the $^1D$ \tr{2s,2s}{2p,2p} genuine doubly-excited states of beryllium is accurately described by all methods considered here.
For this reason, beryllium is discarded in the evaluation of the global statistics discussed in Sec.~\ref{sec:stat}.
Our safe TBE of \SI{7.151}{\eV} matches the previous one, \cite{Veril_2021}
and is consistent with the \SI{7.059}{\eV} value obtained from explicitly correlated calculations, \cite{Galvez_2002}
which remains the most accurate theoretical estimate for this transition.

The two genuine doubly-excited states of carbon dimer (\ce{C2}), $^1\Delta_g$ and $^1\Sigma^+_g$, both of \tr{\pi,\pi}{\sigma^\star,\sigma^\star} character,
are theoretically very challenging for a 12-electron system.
From the well-converged CIPSI calculations, we report safe TBE values of
\SI{2.091}{\eV} ($^1\Delta_g$) and \SI{2.420}{\eV} ($^1\Sigma^+_g$).
While they match the previous values, \cite{Veril_2021} here we reach a level of accuracy of \SI{0.001}{\eV}.
The carbon dimer has an additional $^3\Sigma^-_g$ \tr{\pi,\pi}{\sigma^\star,\sigma^\star} genuine doubly-excited state, not addressed before in the \textsc{quest} database. \cite{Veril_2021}
This genuine double excitation has the lowest excitation energy from the present set.
Our CIPSI/AVTZ calculations provide a safe TBE of \SI{1.258}{\eV}, with an uncertainty of \SI{0.002}{\eV} only.

For the three states of \ce{C2}, we are able to include up to sextuple excitation in CC (CCSDTQPH) in the 6-31+G* basis,
which yields results within the tiny (\SI{0.001}{\eV}) error bars of the exFCI results with the same basis set.
In turn, CCSDTQP overestimates the TBEs minimally, by \SI{0.001}{\eV} to \SI{0.002}{\eV},
whereas CCSDTQ is chemically accurate but has larger errors since it overestimates the TBEs by \SI{0.025}{\eV} on average.

By interpolating the potential energy curves obtained with SHCI in the cc-pV5Z basis from Ref.~\onlinecite{Holmes_2017} to the bond length employed here, we obtain
\SI{2.075}{\eV} ($^1\Delta_g$), \SI{2.412}{\eV} ($^1\Sigma^+_g$), and \SI{1.266}{\eV} ($^3\Sigma^-_g$),
which deviate from our TBEs by less than \SI{0.02}{\eV}.

The carbon trimer (\ce{C3}) is another system with two theoretically challenging genuine doubly-excited states,
of $^1\Delta_g$ and $^1\Sigma^+_g$ symmetries, and of \tr{n,n}{\pi^*,\pi^*} character.
(As first mentioned in Ref.~\onlinecite{Kossoski_2023}, these two transitions were wrongly assigned as \tr{\pi,\pi}{\sigma^*,\sigma^*} in Ref.~\onlinecite{Loos_2019c}.)
We report safe TBEs of \SI{5.230}{\eV} ($^1\Delta_g$) and \SI{5.908}{\eV} ($^1\Sigma^+_g$), matching the previous estimates. \cite{Veril_2021}
They are obtained from CIPSI calculations with associated extrapolation errors between \SI{0.004}{\eV} and \SI{0.017}{\eV}.
Among all the states considered here, these two doubly-excited states of \ce{C3} are arguably the most challenging for the higher-order CC methods.
CC4 and CCSDTQ respectively overestimate the TBEs by \SI{0.16}{\eV} and \SI{0.10}{\eV} on average, and are thus not chemically accurate for these states.
One has to go up to CCSDTQP to approach the exFCI values, which remain overestimated by around \SI{0.01}{\eV}.

We address another state of carbon trimer, a $^1\Pi_g$ \tr{n}{\pi^*} partial doubly-excited state, also not considered in Ref.~\onlinecite{Loos_2019c}.
Our CIPSI/AVTZ calculations directly yield a safe TBE of \SI{4.008}{\eV}.
Since the doubly-excited character is not dominant, CC methods converge much faster than seen for genuine doubles.
We are not aware of previous works for this state to compare with.

\subsection{Nitroxyl and Nitrosomethane}
\label{sec:i2}

Nitroxyl has a low-lying $^1A'$ \tr{n,n}{\pi^*,\pi^*} genuine doubly-excited state.
We report a safe TBE of \SI{4.333}{\eV}, also matching the previous one of \SI{4.33}{\eV}, \cite{Veril_2021}
but here given with an uncertainty associated with the exFCI extrapolation error, of \SI{0.001}{\eV} only.
With the AVDZ basis set, CCSDTQP overestimates exFCI by \SI{0.002}{\eV},
similar to what was observed with the smaller 6-31+G* basis. \cite{Loos_2019c}
Our TBE (\SI{4.333}{\eV}) agrees perfectly with available diffusion Monte Carlo (DMC) results [\SI{4.32(1)}{\eV}]. \cite{Shepard_2022}

An analogous $^1A'$ \tr{n,n}{\pi^*,\pi^*} genuine doubly-excited state is present in nitrosomethane.
The previous TBE of \SI{4.76}{\eV}, obtained from CIPSI calculations, had an error bar of \SI{0.04}{\eV}, making it barely safe. \cite{Loos_2019c}
In the present CIPSI calculation, we are able to considerably reduce the extrapolation error for both the
6-31+G* (from \SI{0.01}{\eV} to \SI{0.002}{\eV}) and AVDZ (from \SI{0.02}{\eV} to \SI{0.012}{\eV}) basis sets,
though not as much for the AVTZ basis (from \SI{0.04}{\eV} to \SI{0.034}{\eV}).
The present TBE is based on the updated exFCI/AVDZ estimate and a CC4 basis set correction.
Our new value, of \SI{4.732}{\eV}, appears \SI{0.028}{\eV} below the previous estimate. \cite{Loos_2019c}
Because the combined extrapolation error of exFCI and the basis set correction error remains below chemical accuracy,
the present TBE of nitrosomethane is considered safe.

\subsection{Ethylene and Formaldehyde}
\label{sec:i3}

A genuine doubly-excited state can be found at relatively high energy in ethylene.
The previously reported TBE for this $^1A_g$ \tr{\pi,\pi}{\pi^*,\pi^*} state was \SI{12.92}{\eV}. \cite{Loos_2019c}
It was obtained from CIPSI calculations, which was assigned as safe despite the relatively large extrapolation error of \SI{0.06}{\eV}.
Here, we update the TBE to \SI{12.899}{\eV}, thanks to new CIPSI calculations in the AVDZ basis (with a small extrapolation error of \SI{0.01}{\eV}) and CCSDTQ basis set correction.
It is also consistent with our exFCI/AVTZ result, of \SI{12.898(29)}{\eV}.
The new TBE is considered safe and appears below the previous estimate by a mere \SI{0.02}{\eV}. \cite{Veril_2021}

Formaldehyde also has a genuine double excitation, of $^1A_1$ symmetry, and \tr{n,n}{\pi^*,\pi^*} character.
The previous TBE, of \SI{10.35}{\eV}, was directly obtained from exFCI results. \cite{Loos_2019c}
Here, we performed new CIPSI calculations and were able to reproduce this value for the fourth state of the $^1A_1$ symmetry, which, however, is a clear singly-excited state.
We find that the doubly-excited state corresponds instead to the fifth state, slightly higher in energy, at \SI{10.425}{\eV}.
The TBE is therefore updated to \SI{10.425}{\eV}, which has an associated uncertainty of \SI{0.019}{\eV} and can thus be considered safe.
The previous misassignment of the correct state explains the large shift of \SI{0.08}{\eV} from the previous to the present TBE.

The doubly-excited state of ethylene displays the largest basis set effects from the present set of systems.
This is not too surprising as the present TBE of \SI{12.899}{\eV} is greater than the vertical ionization threshold,
reported between \SI{10.75}{\eV} and \SI{10.95}{\eV}. \cite{Holland_1997,Davidson_1998,Musia_2004}
The state is therefore a resonance, but its interaction with the ionization continuum is neglected in our calculations.
It is also worth mentioning that increasing the basis set from 6-31+G* to AVDZ has a large effect on the doubly-excited state of formaldehyde.
Despite being close to, our TBE (\SI{10.425}{\eV}) appears below the vertical ionization energy,
estimated experimentally at \SI{10.9}{\eV}, \cite{VonNiessen_1980}
and obtained with exFCI calculations in the same basis and geometry as ours at \SI{10.90}{\eV}. \cite{Marie_2024}
For the multiconfigurational methods, the largest basis set errors are found for ethylene and formaldehyde, as further discussed in Sec.~\ref{sec:stat}.

\subsection{Butadiene, Glyoxal, and Acrolein}
\label{sec:i4}

Butadiene has a well-known $^1A_g$ \tr{\pi}{\pi^*} state with a significant doubly-excited character.
The former TBE of \SI{6.50}{\eV} was based on the exFCI/AVDZ result and a CCSDT basis set correction.
Here, while we adopt the same AVTZ basis set correction, we change the protocol for the first term.
Indeed, we rely on the CCSDTQ/Pop result combined with a AVDZ basis set correction obtained at the CC4 level,
which should have a significantly smaller uncertainty than the extrapolation error of exFCI/AVDZ.
The present TBE of \SI{6.515}{\eV} is considered safe, and blueshifts the previous one by \SI{0.015}{\eV} only. \cite{Veril_2021}

Glyoxal has a genuine $^1A_g$ doubly-excited state of \tr{n,n}{\pi^*,\pi^*} character, which is relatively low-lying in energy.
The previous exFCI/Pop and exFCI/AVDZ, of \SI{5.60}{\eV} and \SI{5.48}{\eV}, had a reported error bar not higher than \SI{0.01}{\eV}, \cite{Loos_2019c} which was probably underestimated.
The latter value, supplemented with a CCSDT basis set correction, yielded a TBE of \SI{5.61}{\eV}, which was considered to be safe due to the small extrapolation errors. \cite{Veril_2021}
We performed new CIPSI calculations with the 6-31+G* and AVDZ basis sets and found considerably larger extrapolation errors, of around \SI{0.04}{\eV} and \SI{0.07}{\eV}, respectively.
To establish a new TBE, we employ the exFCI/Pop result, followed by a AVDZ basis set correction with CC4, and then by a AVTZ basis set correction with CASPT3.
\fk{Alternatively, we could have chosen CCSDT for the basis set correction. However,}
as discussed in more detail in Sec.~\ref{sec:stat}, CCSDT (as well as CC3) are not reliable methods to correct for basis set effects in the case of genuine doubly-excited states.
Our new protocol yields \fk{\SI{5.492}{\eV}} for the TBE of glyoxal,
\fk{which is deemed unsafe due to the extrapolation error of exFCI/Pop (\fk{\SI{0.034}{\eV}}) and the additional basis set correction error. This is a borderline case and the uncertainty for our TBE is probably below \SI{0.1}{\eV}.}
The revised TBE significantly redshifts (\fk{\SI{-0.12}{\eV}}) the previous estimate of \SI{5.61}{\eV}, \cite{Veril_2021}
mostly reflecting our more reliable basis set correction. It also underestimates (\fk{\SI{-0.138}{\eV}}) the available DMC result of \SI{5.63(1)}{\eV}. \cite{Shepard_2022}

The $^1A'$ \tr{\pi}{\pi^*} state of acrolein has a partial doubly-excited character.
The first TBE reported for this state, of \SI{7.87}{\eV}, was obtained from exFCI/Pop results and a CC3 basis set correction. \cite{Loos_2019c,Veril_2021}
It was later revised to \SI{7.929}{\eV}, from CC4/AVDZ and CCSDT/AVTZ basis set corrections. \cite{Loos_2022b}
Here, we were able to perform CCSDTQ/Pop calculations, which yield an excitation energy only \SI{0.001}{\eV} below the CC4 result in the same basis.
Using the same protocol for the basis set corrections, the new TBE is essentially unchanged at \SI{7.928}{\eV}, and is considered to be safe.

\subsection{Cyclopentadiene, Benzene, and Hexatriene}
\label{sec:i5}

The \tr{\pi}{\pi^*} $^1A_1$ partial doubly-excited state ($\%T_1 = 79\%$) of cyclopentadiene was not considered in Ref.~\onlinecite{Loos_2019c},
but was later addressed and extensively discussed in Refs.~\onlinecite{Loos_2020c} and~\onlinecite{Loos_2022b}.
In the former contribution, \cite{Loos_2020c} a TBE of \SI{6.523}{\eV} was reported, based on CCSDT/AVTZ calculations.
The latter publication \cite{Loos_2022b} updated it to a safe value of \SI{6.452}{\eV}, based on the CC4/Pop result and CCSDT basis set corrections,
a significant reduction of \SI{0.07}{\eV} with respect to the former.
Here, we were able to perform both CCSDTQ/Pop and CC4/AVDZ calculations. Combining these results with the CCSDT basis set correction,
the TBE of cyclopentadiene moves negligibly, to \SI{6.451}{\eV}. This is a safe estimate.

Benzene has a \tr{\pi}{\pi^*} $^1E_{2g}$ partial doubly-excited state ($\%T_1 = 73 \%$).
The previous TBE, of \SI{8.28}{\eV}, was determined using the exFCI/Pop result and a CC3 basis set correction. \cite{Loos_2019c,Veril_2021}
It was considered to be safe because of the small extrapolation error of exFCI.
Here, we performed new CIPSI calculations but could not locate this state with a sufficiently small uncertainty,
which makes us believe that the previous extrapolation error was optimistically too small. \cite{Loos_2019c}
The present CC4/AVDZ calculations and CCSDT basis set correction allow us to revise the TBE to \SI{8.190}{\eV},
which is below the previous value by \SI{0.09}{\eV}.
Due to the sizeable gap of \SI{0.18}{\eV} between CC4/AVDZ and CCSDT/AVDZ and the significant double component ($\%T_1 = 73 \%$), we now assign the TBE of benzene as unsafe.

Benzene also has a \tr{\pi,\pi}{\pi^*,\pi^*} $^1A_{1g}$ genuine doubly-excited state,
here described with multiconfigurational methods only, since high-order CC or CIPSI calculations cannot clearly identify it.
CASPT3 yields excitation energies in-between the CASTP2(IPEA) and SC-NEVPT2 results, whereas PC-NEVPT2 appears to underestimate the true value.
Our unsafe TBE of \SI{10.315}{\eV} is thus based on CASPT3/AVTZ,
which shifts the previous and also unsafe estimate, \cite{Veril_2021} based on extended multistate (XMS-)CASPT2, by \SI{-0.235}{\eV}.
Similarly to ethylene, the genuine doubly-excited state of benzene lies above its vertical ionization threshold, theoretically reported at 9.448 eV. \cite{Deleuze_2003}

Hexatriene has a $^1A_g$ \tr{\pi}{\pi^*} excitation with a significant partial doubly-excited character ($\%T_1 = 65 \%$).
The first and unsafe estimate of its TBE, reported at \SI{5.619}{\eV}, was based on CCSDT/AVDZ results and a basis set correction at the CC3 level of theory. \cite{Veril_2021}
It was later revised to \SI{5.459}{\eV}, which was considered a safe estimate (though an upper limit), as obtained from a CC4/Pop calculation plus a CC3 basis set correction. \cite{Loos_2022b}
The present CCSDT results indicate a AVTZ $-$ Pop basis set effect of a different sign than the one obtained with CC3, suggesting that neither of them is a reliable option for hexatriene.
Here, instead, we employ CASPT3 for the basis set correction, on top of the CC4/Pop value,
which yields a new TBE of \SI{5.435}{\eV}, shifted by \SI{-0.02}{\eV} with respect to the former value. \cite{Loos_2022b}
Our revised TBE is also consistent with a second estimate from Ref.~\onlinecite{Loos_2022b}, of \SI{5.43}{\eV}, which was put forward based on a comparison between results of hexatriene and butadiene.
Although its uncertainty is likely below \SI{0.05}{\eV}, we prefer to be conservative and assign the present TBE of hexatriene as unsafe.

\subsection{Pyrazine and Tetrazine}
\label{sec:i6}

We address two close-lying states with doubly-excited character of pyrazine,
a $^1A_g$ \tr{n}{\pi^*} partial double and a $^1A_g$ \tr{n,n}{\pi^*,\pi^*} genuine double.
The previous unsafe TBE (\SI{8.037}{\eV}) for the latter was the PC-NEVPT2/AVTZ estimate. \cite{Veril_2021}
The present CC4 calculations yield excitation energies significantly below PC-NEVPT2, whereas CASPT3 appears closer to CC4 than PC-NEVPT2 does.
For these reasons, here we rely on the CC4/AVDZ result and a CASPT3 basis set correction to provide an
improved TBE for the genuine double of pyrazine at \SI{7.904}{\eV}.
The new value displaces the previous estimate by \SI{-0.13}{\eV}, yet remains unsafe.
We notice that the impact of increasing the basis set from 6-31+G* to AVDZ differs substantially with
CC4 (\SI{-0.06}{\eV}) and multiconfigurational methods (\SIrange{-0.13}{-0.12}{\eV}).
In addition, the partial and genuine doubly-excited states are strongly mixed at some levels of CC, which could cause unusual basis set effects.
The present TBE is therefore subject to significant uncertainty.

The previously reported TBE for the $^1A_g$ \tr{n}{\pi^*} partial doubly-excited state of pyrazine
stemmed from CC3/AVTZ calculations. \cite{Veril_2021}
The value of \SI{8.697}{\eV} was labeled unsafe because of its significant partial doubly-excited character ($\%T_1 = 71\%$).
Here, we rely on CC4/AVDZ calculations and basis set correction from CASPT3 to improve the TBE to \SI{8.480}{\eV},
lying, as expected, significantly below (\SI{-0.22}{\eV}) the previous estimate.
While likely significantly more accurate than the previous TBE, we still assign it as being unsafe.
Had we used CCSDT for the basis set correction,
or instead a CASPT3 calculation with no Rydberg orbitals in the active space, the TBE would change very little,
by only \SI{0.002}{\eV} and \SI{0.006}{\eV}, respectively, hinting that the error associated with the basis set correction is probably small.

Tetrazine has three genuine \tr{n,n}{\pi^*,\pi^*} doubly-excited states.
The lowest-lying $^1A_g$ state is formed by
exciting two electrons from the same non-bonding orbital into the same $\pi^*$ orbital,
whereas in the $^3B_{3g}$ and $^1B_{3g}$ states, each electron is promoted to a different $\pi^*$ orbital.
The previous TBEs, of \SI{4.608}{\eV} ($^1A_g$), \SI{5.506}{\eV} ($^3B_{3g}$), and \SI{6.149}{\eV} ($^1B_{3g}$), were obtained from PC-NEVPT2 calculations, \cite{Loos_2020c}
and were estimated to have an error bar of \SI{\pm 0.1}{\eV} (thus unsafe). \cite{Loos_2022b}

Starting with the $^1A_g$ state in the 6-31+G* basis, we find consistent excitation energies among CASPT3 (\SI{5.03}{\eV}), CC4 (\SI{5.06}{\eV}), and exFCI [\fk{\SI{4.99(7)}{\eV}}],
whereas PC-NEVPT2 provides a significantly lower value (\SI{4.75}{\eV}), close to the CASPT2 one (\SI{4.72}{\eV}).
Although the \SI{0.07}{\eV} extrapolation error of the present exFCI is not small enough to define a safe TBE, it allows us to confidently rule out PC-NEVPT2 and SC-NEVPT2 as safe choices.
Similar trends are observed with the AVDZ basis set, even though we do not have an exFCI estimate in this basis.
From the overall consistency among the CASPT3, CC4, and exFCI, and since we have results for the three bases with CASPT3, this is the method of choice to provide the TBE for the $^1A_g$ state.
The same reasoning holds for both the $^3B_{3g}$ and $^1B_{3g}$ doubly-excited states,
and we therefore also rely on CASPT3 to provide the TBEs for these states.
For the $^1B_{3g}$ state, we notice that PC-NEVPT2 performs better than for the other states, although still underestimating the excitation energy.

The new TBEs, of \SI{4.951}{\eV} ($^1A_g$), \SI{5.848}{\eV} ($^3B_{3g}$), and \SI{6.215}{\eV} ($^1B_{3g}$), are certainly more reliable than the previous values,
which change substantially, by \SI{+0.34}{\eV}, \SI{+0.34}{\eV}, and \SI{+0.07}{\eV}, respectively.
Although the revised TBEs of tetrazine are still considered to be unsafe, the overall consistency between the various methodologies
indicate that the errors are small, probably of the order of \SI{0.1}{\eV} or less.
We also underline that the present TBE for the $^1A_g$ doubly-excited state, of \SI{4.951}{\eV}, is reasonably close to the available DMC value
of \SI{4.99(1)}{\eV}. \cite{Shepard_2022}

\section{Double Excitations: Extension}
\label{sec:Extension}

\subsection{Criegee's Intermediate}
\label{sec:e1}

Here, we address the first two singlet excited states of the simplest Criegee's intermediate (\ce{CH2OO}),
both having a partial doubly-excited character ($\%T_1 = 80 \%$).
For the lowest-lying \fk{$^1A''$} \tr{n}{\pi^*} state, we report a safe TBE of \SI{2.403}{\eV},
based on the CIPSI/AVDZ result and a CC4 basis set correction.
In comparison to the most recent theoretical data,
our TBE is substantially lower than MRCI/AVTZ (\SI{2.70}{\eV}) \cite{Meng_2014}
and closer to CASPT2/AVDZ (\SI{2.43}{\eV}) \cite{Esposito_2021} results.
The latter agrees well with our own CASPT2/AVDZ value of \SI{2.371}{\eV}.

The second singlet state, $^1A'$ \tr{\pi}{\pi^*}, is optically bright and has received significant attention due to its involvement in atmospheric chemistry. \cite{Beames_2012,Ting_2014,Meng_2014,Dawes_2015}
Previous calculations find this state at \SI{3.89}{\eV} (with MRCI/AVTZ) \cite{Meng_2014} and at \SI{3.74}{\eV} (with CASPT2/AVDZ). \cite{Esposito_2021}
Our own CASPT2/AVDZ result yields \SI{3.932}{\eV}.
The most accurate available calculation was performed by Dawes et al., \cite{Dawes_2015}
who reported an excitation energy of \SI{3.751}{\eV} obtained at the MRCI-F12/cc-pVTZ level of theory.

Here, we employ CIPSI/AVDZ and a CC4 basis set correction to provide a safe TBE of \SI{3.715}{\eV} for the $^1A'$ partial double of this Criegee's intermediate.
Our value deviates only by \SI{-0.036}{\eV} from the result of Ref.~\onlinecite{Dawes_2015}.
The present TBE of \SI{3.715}{\eV} also fits the measured photoabsorption spectra, \cite{Beames_2012,Ting_2014} which reported the maximum of the absorption band at \SI{335}{\nm} (\SI{3.70}{\eV}) \cite{Beames_2012} and at \SI{340}{\nm} (\SI{3.65}{\eV}), \cite{Ting_2014} though we recall that such comparisons between vertical transition energies and measured peak maxima always come with significant approximations.

\subsection{Nitrous acid}
\label{sec:e2}

Despite its small size, nitrous acid is a challenging system that deserves a dedicated discussion.
We are interested in its $^1A'$ \tr{n,n}{\pi^*,\pi^*} genuine doubly-excited state,
which appears as the third and/or fourth state in the $^1A'$ symmetry, and with a more genuine or partial character, depending on the level of theory and basis set.

In the 6-31+G* basis set, there is a strong mixing between the \tr{n,n}{\pi^*,\pi^*} doubly-excited and a \tr{n}{3s} Rydberg singly-excited configurations,
giving rise to two close-lying adiabatic states (third and fourth in the $^1A'$ symmetry).
This is observed in our well-converged CIPSI calculations, and at the CCSDTQ and multiconfigurational levels alike, and is thus taken as the qualitatively correct picture for this basis.
In this sense, there is no single adiabatic state with a well-defined genuine doubly-excited state, but rather two states with a doubly-excited configuration,
combined in-phase or out-of-phase with a Rydberg configuration.
If the $3s$ Rydberg orbital is not included in the active space of multiconfigurational calculations, a single genuine double excitation is described.
Similarly, by downgrading to CC4 and CCSDT, the two configurations are no longer strongly mixed, and one can find a genuine doubly-excited state and a Rydberg singly-excited state.
Surprisingly, the mixing returns at the CC3 level, but for the wrong reason, given the opposite trend observed in CCSDT and CC4.
In fact, at the CC3 level, the doubly-excited configuration mixes with a different, higher-lying configuration.

Clearly, the close proximity and strong coupling between the two states prove to be very challenging for CC methods.
Once the CCSDTQ level is reached, the excitation energies
(\SI{7.979}{\eV} and \SI{8.185}{\eV})
are in very good agreement with our CIPSI results
[\SI{7.979(40)}{\eV} and \SI{8.170(41)}{\eV}].
The singly- and doubly-excited configurations are important for both states, which are combined in-phase for the lower-lying state, and out-of-phase for the higher-lying one.
The mixing is weak at the CASSCF level but becomes pronounced after the perturbative treatment.

Augmenting the basis sets to AVDZ or to AVTZ, the \tr{n}{3s} Rydberg configuration is largely stabilized in comparison to the \tr{n,n}{\pi^*,\pi^*} one,
which suppresses the strong mixing observed in the small basis.
Indeed, 6-31+G* is too compact to properly model the Rydberg configuration, which appears too high in energy and artificially mixes with the doubly-excited configuration.
In the AVDZ and AVTZ bases, the fourth $^1A'$ state can be unambiguously assigned as the genuine double excitation across all methods,
despite the relatively high value of $\%T_1$ (50\%) obtained at the CC3 level.
The multiconfigurational methods are overall consistent among themselves, with the CASPT3 energy virtually matching the \fk{CIPSI} one in the AVDZ basis,
while the NEVPT2 values are overestimated by around \SI{0.1}{\eV}.
In the AVDZ basis, CCSDTQ yields an excitation energy (\SI{8.081}{\eV}) within the CIPSI error bar [\fk{\SI{8.057(32)}{\eV}}].
We notice a substantial basis set effect \fk{from AVDZ to AVTZ}, which decreases the excitation energies by around \fk{\SI{0.1}{\eV}}.
\fk{Here, we take the CIPSI/AVTZ result to obtain a safe TBE of \SI{7.969}{\eV}, which has an estimated error of \SI{0.036}{\eV}.}
To the very best of our knowledge, there is no available experimental or theoretical data on the doubly-excited state of nitrous acid.

\subsection{Cyclobutadiene and Diazete}
\label{sec:e3}

The $^1A_g$ \tr{\pi,\pi}{\pi^*,\pi^*} genuine doubly-excited state of cyclobutadiene has been extensively discussed in Ref.~\onlinecite{Monino_2022a}.
There, the authors provide a TBE of \SI{4.038}{\eV}.
Here, we employ CASPT3 results to provide the AVTZ basis set correction, instead of PC-NEVPT2 used in Ref.~\onlinecite{Monino_2022a}.
This change is motivated by the very close agreement between CASPT3 and the TBE/Pop and TBE/AVDZ values.
The updated and safe TBE shifts minimally (\SI{-0.002}{\eV}), to \SI{4.036}{\eV}.

The diamond-shaped diazete has a $^1A_1$ genuine doubly-excited state of \tr{\pi,\pi}{\pi^*,\pi^*} character, analogous to the one of cyclobutadiene.
To the best of our knowledge, neither this nor other excited states have been previously investigated using high-level wave function approaches.
We employ the same composite method as for cyclobutadiene to provide a TBE of \SI{6.605}{\eV} for the genuine double excitation of diazete.
We believe this is probably a safe estimate due to the small energy gap between CC4/Pop and CCSDTQ/Pop, of \SI{0.004}{\eV},
the fact that CCSDTQ/Pop yields an excitation energy (\SI{6.726}{\eV}) in excellent agreement with CIPSI, at \SI{6.727(47)}{\eV},
and the overall parallel between diazete and the safe case of cyclobutadiene.

\subsection{Borole}
\label{sec:e4}

We address two excited states of borole with significant double contributions,
a lower-lying genuine $^1A_1$ \tr{\pi,\pi}{\pi^*,\pi^*} state,
and a higher-lying $^1A_1$ \tr{\pi}{\pi^*} state with partial double character.

For the latter, we provide a TBE of \SI{6.484}{\eV}.
This is based on the CCSDTQ/Pop result and basis set corrections done at the CC4/AVDZ and CCSDT/AVTZ levels of theory.
The CC series rapidly converges in the 6-31+G* basis, while basis set effects are consistent across various methods.
Further taking into account the relatively high value of $\%T_1$ (81\%),
we conclude that our TBE is safe for this partial double excitation of borole.

Similarly, based on CCSDTQ/Pop with additional CC4 and CASPT3 basis set corrections for AVDZ and AVTZ, respectively,
we provide a TBE of \SI{4.708}{\eV} for the genuine doubly-excited state.
This estimate is assigned as unsafe, though, as we could not reduce the CIPSI/Pop incertitude below \fk{\SI{0.049}{\eV}}.

We have not found any previous studies regarding the excited states of borole,
except for a recent work of our groups that evaluated state-specific CC methods to describe its genuine double excitation. \cite{Damour_2024}

\subsection{Cyclopentadienone and Cyclopentadienethione}
\label{sec:e5}

Cyclopentadienone has four low-lying excited states with relevant double characters.
Here, we investigate the $^1A_1$ \tr{\pi}{\pi^*} partial double and the $^1A_1$ \tr{\pi,\pi}{\pi^*,\pi^*} genuine double,
in addition to the $^1B_1$ \tr{n,\pi}{\pi^*,\pi^*} and $^3B_1$ \tr{n,\pi}{\pi^*,\pi^*} genuine doubles.

Starting with the latter two, we provide TBEs based on CIPSI/Pop and CASPT3 basis set corrections.
The present values, \SI{5.009}{\eV} ($^1B_1$) and \SI{4.821}{\eV} ($^3B_1$),
change by \SI{-0.016}{\eV} ($^1B_1$) and \SI{-0.092}{\eV} ($^3B_1$) with respect to the previous unsafe estimate, which relied on PC-NEVPT2/AVTZ. \cite{Veril_2021}
Although more accurate, the present TBEs are still deemed unsafe, because of the sizeable CIPSI extrapolation errors (\SI{0.02}{\eV} and \SI{0.04}{\eV}) and
the additional CASPT3 basis set error, which averages at \SI{0.015}{\eV} for the present set (see the \SupInf).
These are borderline cases though, and the present errors are probably below \SI{0.1}{\eV}.

Due to their strong mixing, the two $^1A_1$ excited states of cyclopentadienone are perhaps the most challenging systems from the present set.
As discussed in Ref.~\onlinecite{Shepard_2022} and also confirmed by our calculations,
both \tr{\pi,\pi}{\pi^*,\pi^*} doubly-excited and \tr{\pi}{\pi^*} singly-excited configurations are important for the two states,
with a dominance of the doubly-excited one in the lower-lying state and of the singly-excited one in the higher-lying one.
It is thus reasonable to label the first as the genuine double and the second as the partial double, and an analysis of the CC4 data supports such an assignment.
This is in contrast to what the CC3 $\%T_1$ values suggest,
based on a lower value ($\%T_1 = 50 \%$) for the partial double and a higher one ($\%T_1 = 74 \%$) for the genuine double.

For the $^1A_1$ genuine, we rely on our CC4/Pop result and CASPT3 basis set corrections to obtain a new yet still unsafe TBE of \SI{5.795}{\eV},
which significantly changes (\SI{-0.292}{\eV}) the previous unsafe estimate from PC-NEVPT2/AVTZ. \cite{Veril_2021}
Our value (\SI{5.795}{\eV}) is somewhat below the available DMC result [\SI{5.90(1)}{\eV}], \cite{Shepard_2022}
deviating from it by \SI{-0.105}{\eV}.

To revise the TBE for the partial double, we also employ the CC4/Pop result, but in combination with a PC-NEVPT2 basis set correction, as CASPT3 yields too high excitation energies for this state.
Our unsafe value of \SI{6.714}{\eV} is redshifted (\SI{-0.052}{\eV}) with respect to the previous and also unsafe value, which was based on CCSDT/AVDZ and a CC3 basis set correction. \cite{Veril_2021}
This brings our TBE further lower in energy in comparison to the DMC result [\SI{6.89(1)}{\eV}], \cite{Shepard_2022} by \SI{-0.176}{\eV}.

Cyclopentadienethione also presents four doubly-excited states analogous to those of cyclopentadienone.
For the $^1B_1$ \tr{n,\pi}{\pi^*,\pi^*} and $^3B_1$ \tr{n,\pi}{\pi^*,\pi^*} genuine doubly-excited states,
we provide TBEs of \SI{3.156}{\eV} and \SI{3.115}{\eV}, respectively, based on the exFCI/Pop results and CASPT3 basis set corrections.
In spite of the small exFCI extrapolation errors, around \SI{0.02}{\eV} and \SI{0.03}{\eV}, we prefer to be conservative here and label these TBEs as unsafe.
It is worth noticing the overall consistency among CASPT3, CASPT2(IPEA), and the two variants of NEVPT2 for these two transitions.

We further address two $^1A_1$ states of cyclopentadienethione with strongly mixed \tr{\pi}{\pi^*} and \tr{\pi,\pi}{\pi^*,\pi^*} characters.
The singly-excited configuration dominates in the lower-lying state (the partial double), whereas the doubly-excited configuration is more pronounced in the higher-lying one (the genuine double).
This is the opposite of what we find for cyclopentadienone.
Here, we update the TBE for the partial double to \SI{5.329}{\eV}, thus changing the previous value by \SI{-0.099}{\eV}. \cite{Veril_2021}
For that, we use CC4/Pop and a CASPT3 basis set correction, while PC-NEVPT2/AVTZ was used in Ref.~\onlinecite{Veril_2021}.

The higher-lying $^1A_1$ genuine double has not yet been considered in the \textsc{quest} database.
Excluding the less reliable CASPT2 result, the other multiconfigurational methods display excitation energies spanning a wide range of approximately \SI{0.7}{\eV}.
The CC4/Pop calculation yields \SI{5.780}{\eV}, which is expected to be slightly overestimating the true value for this basis set.
For this genuine double, we employ CC4/Pop and the PC-NEVPT2 basis set correction to provide a TBE of \SI{5.555}{\eV}.

\fk{For the higer-lying $A_1$ states of cyclopentadienone and cyclopentadienethione, we also tested extended multistate CASPT2 (XMS-CASPT2) variants. \cite{Granovsky_2011,Shiozaki_2011}
As discussed above, these states strongly mix with a lower-lying $A_1$ state, which explains the large spread of vertical transition energies observed between the different multiconfigurational methods.
At the XMS-CASPT2/AVTZ level, the excitation energies decrease when compared to CASPT2(IPEA)/AVTZ and approach the TBEs.
For cyclopentadienone, we obtain \SI{7.172}{\eV} [CASPT2(IPEA)], \SI{6.547}{\eV} (XMS-CASPT2), and \SI{6.714}{\eV} (TBE), whereas for cyclopentadienethione the corresponding values are
\SI{5.679}{\eV} [CASPT2(IPEA)], \SI{5.425}{\eV} (XMS-CASPT2), and \SI{5.555}{\eV} (TBE).}

\subsection{Oxalyl fluoride}
\label{sec:e6}

Oxalyl fluoride is structurally similar to glyoxal, with hydrogen atoms replaced by fluorines.
In close analogy to glyoxal, oxalyl fluoride has a $^1A_g$ \tr{n,n}{\pi^*,\pi^*} genuine doubly-excited state,
though much higher in energy due to the inductive effect of the halogen atoms.

All our multiconfigurational calculations yield comparable excitation energies,
with the exception of CASPT2(IPEA), too low by \SI{0.1}{\eV}, and CASPT2 which is not very reliable as expected.
CASPT3, CASPT3(IPEA), PC-NEVPT2, and SC-NEVPT2 results differ among themselves by no more than \SI{0.046}{\eV}, \SI{0.060}{\eV}, and \SI{0.035}{\eV}, in the 6-31+G*, AVDZ, and AVTZ basis sets.
Meanwhile, CC4 yields a value higher than the multiconfigurational methods by \SI{0.15}{\eV} on average, in both 6-31+G* and AVDZ basis.
This is a typical error of CC4 for genuine doubles, albeit in the higher range of the expected error bar, as discussed in Sec.~\ref{sec:stat}.
We thus rely on CASPT3 to provide the TBE, which yields energies slightly above CASPT3(IPEA) and slightly below the two NEVPT2 variants.
CASPT3/AVTZ calculation provides a TBE of \SI{8.923}{\eV} for the genuine doubly-excited state of oxalyl fluoride.
In the absence of more definitive results, we prefer to assign this TBE as unsafe, even though the overall consistency among the different methodologies suggests a reasonable estimate.

Besides our recent investigation that targeted this genuine double excitation with state-specific CC methods, \cite{Damour_2024}
we are not aware of other studies for this compound.

\subsection{Benzoquinone, Octatetraene, and Naphthalene}
\label{sec:e7}

The excited states of benzoquinone have been addressed before in the \textsc{quest} database, \cite{Veril_2021}
but they have not been extensively discussed.
Here, we revisit six states having a relevant doubly-excited component: the genuine $^1A_g$ \tr{n,n}{\pi^*,\pi^*} state,
the partial $^1A_g$ \tr{\pi}{\pi^*}, and the partial $^1B_{3u}$, $^1B_{2g}$, $^1A_u$, and $^1B_{1g}$ states, the latter four having a dominant \tr{n}{\pi^*} character.

For the $^1A_g$ genuine double, our CC4/Pop result is slightly below CASPT3 and above PC-NEVPT2,
which suggests the latter method is probably closer to the true value.
The unsafe TBE reported in Ref.~\onlinecite{Veril_2021}, obtained from PC-NEVPT2/AVTZ calculations, thus remains unchanged, at \SI{4.566}{\eV}.
For the $^1A_g$ \tr{\pi}{\pi^*} and the $^1B_{1g}$ \tr{n}{\pi^*} states presenting a partial double character, we adopt CC4/Pop, a AVDZ basis set correction with CCSDT, and a AVTZ basis set correction with CC3,
to yield TBEs of \SI{6.351}{\eV} ($^1A_g$) and \SI{6.469}{\eV} ($^1B_{1g}$). Due to their relatively small $\%T_1$ (64\% and 70\%), these TBEs are unsafe.
For the three remaining partial doubles, the previous TBEs were based on CCSDT/AVDZ and a CC3 basis set correction. \cite{Veril_2021}
Here, we revise them with CC4/Pop results plus a AVDZ basis set correction with CCSDT and a AVTZ basis set correction with CC3, i.e., we use
the same approach for all partial doubly-excited states. This protocol yields \SI{5.656}{\eV} ($^1B_{3u}$), \SI{5.764}{\eV} ($^1B_{2g}$), and \SI{6.083}{\eV} ($^1A_u$), all considered to be safe,
given the relatively large $\%T_1$ (80\%, 76\%, and 75\%, respectively). They are significantly below the previous estimates, \cite{Veril_2021} by \SI{-0.14}{\eV}, \SI{-0.21}{\eV}, and \SI{-0.26}{\eV}.
The shift is attributed largely to the quadruple excitations of CC4, proven to be significant even for states with significant values of $\%T_1$ (from 75\% to 80\%), consistently with
the findings of Ref.~\onlinecite{Loos_2022b}.

The excited states of benzoquinone have been the subject of numerous theoretical studies. \cite{Pou-Amerigo_1999,Weber_2001,Schreiber_2008,Bousquet_2013,Jones_2017}
Our TBE for the $^1B_{3u}$ state (\SI{5.656}{\eV}) is comparable to the reference value of \SI{5.60}{\eV} recommended in Ref.~\onlinecite{Schreiber_2008},
based on CASPT2(IPEA)/TZVP calculations. Our own CASPT2(IPEA) calculations in the AVTZ basis yield \SI{5.774}{\eV}.
Here, we do not perform an exhaustive comparison for the other states, which have been described with more approximate
methods before, like CASPT2 without IPEA shift \cite{Pou-Amerigo_1999,Weber_2001} and TD-DFT. \cite{Jones_2017}
Transitions to the states of benzoquinone addressed here have not been spectroscopically identified, \cite{Jones_2017}
as they are optically dark or have very low oscillator strengths. \cite{Pou-Amerigo_1999,Weber_2001,Schreiber_2008,Bousquet_2013,Jones_2017}

In parallel to butadiene and hexatriene, octatetraene also has a
$^1A_g$ \tr{\pi}{\pi^*} state with partial doubly-excited character.
It appears lower in energy than the analogous state of hexatriene, which in turn is lower than in butadiene.
Octatetraene is the smallest polyene where the dark $^1A_g$ state is lower lying than the bright $^1B_u$ \tr{\pi}{\pi^*} singly-excited state. \cite{Nakayama_1998,Schreiber_2008}
The previous unsafe TBE for the $^1A_g$ state, of \SI{4.901}{\eV}, relied on the CCSDT/Pop result and a CC3 basis set correction. \cite{Veril_2021}
Because of its partial double character ($\%T_1 = 64\%$), this previous estimate should be too large, appearing above the TBE for the $^1B_u$ state, reported at \SI{4.78}{\eV} following the same protocol. \cite{Veril_2021}
Here, we opt for CASPT3/AVTZ to provide an updated TBE of \SI{4.680}{\eV} \fk{for the $^1A_g$ state}, substantially redshifted (\SI{-0.221}{\eV}) with respect to the previous one.
While probably more accurate, it remains an unsafe estimate.
With this revised value, we obtain the correct ordering of $^1A_g$ and $^1B_u$ states.
Our TBE is very close to the reference value of \SI{4.66}{\eV} chosen by Thiel and coworkers, \cite{Schreiber_2008}
obtained from multireference M{\o}ller-Plesset perturbation theory and basis set extrapolation. \cite{Nakayama_1998}
\fk{It also matches the internally-contracted multireference CCSD (icMRCCSD) result in the TZVP basis set reported in Ref.~\onlinecite{Zielinski_2023} (\SI{4.68}{\eV}).
It is worth mentioning the equally remarkable agreement for hexatriene (\SI{5.44}{\eV} from icMRCCSD/TZVP \cite{Zielinski_2023} compared with our TBE of \SI{5.435}{\eV})
and butadiene (\SI{6.52}{\eV} from icMRCCSD/TZVP \cite{Zielinski_2023} compared with our TBE of \SI{6.515}{\eV}).
We notice, however, that the geometries employed in Ref.~\onlinecite{Zielinski_2023} are not the same as ours.}

Naphthalene has a $^1A_g$ \tr{\pi}{\pi^*} state with partial double character.
This is the last and the largest system in the present set.
CCSDT probably overestimates the real excitation energy, whereas PC-NEVPT2 yields comparable results, and CASPT3 yields somewhat lower values.
For this reason, our reported TBE of \SI{6.748}{\eV} results from the CASPT3/AVTZ calculation.
The new value, considered unsafe, is below the previous estimate by \SI{-0.126}{\eV},
which was based on CCSDT/Pop and CC3 basis set correction. \cite{Veril_2021}

Our revised TBE is closer to the reference value reported in Ref.~\onlinecite{Schreiber_2008}
(\SI{6.71}{\eV}), which was based on CASPT2(IPEA) calculations in the TZVP basis set.
The latter value is consistent with our own CASPT2(IPEA)/AVTZ result of \SI{6.793}{\eV}, the main difference being the presence of diffuse functions in our case.
For a comprehensive comparison with more approximate theoretical results on naphthalene, we refer the interested reader to Refs.~\onlinecite{Fliegl_2014} and~\onlinecite{Sauri_2011}.
Here, we highlight that the inclusion of a second set of diffuse basis functions at the CC2 level of theory was found to have a significant stabilization effect (\SI{-0.21}{\eV}) for the $^1A_g$ excitation energy
even though this is a valence state. \cite{Fliegl_2014}
It is not clear at this point if such a pronounced effect also occurs at a higher level of theory, or if it is an artifact of CC2.
Finally, our TBE (\SI{6.748}{\eV}) significantly overestimates the reported value from two-photon spectroscopy measurements, at \SI{6.05}{\eV}. \cite{Dick_1981}

\section{Statistics}
\label{sec:stat}

Once established the TBEs for the vertical excitation energy of each state and for the three basis sets, we computed their energy differences
with respect to the CC and multireference results.
Table~\ref{tab:errors} gathers the usual statistical measures, for the ensemble of states and is also separated by partial and genuine doubly-excited states.
For this purpose, we take into account the TBEs for the three bases, though limited to those labeled as safe.
To obtain the statistics for a given method, we do not discard the states where such a method is used to define the TBE.
We do mention below, however, the effect of removing such states from the statistics, which can be relevant for CCSDTQ and CC4.
The underlying distribution of errors, for both safe and unsafe cases, is presented in the {\SupInf}.

\begin{table}[ht!]
\caption{Mean signed error (MSE), mean absolute error (MAE), root-mean-square error (RMSE), and standard deviation of the error (SDE), in units of eV, with respect to the TBEs,
for various CC and multiconfigurational methods,
including all excitations labeled as safe, and the subsets of genuine and partial doubly-excited states, while accounting for the three basis sets.}
\label{tab:errors}
\begin{ruledtabular}
\begin{tabular}{lddddd}
Method            & \mc{1}{c}{\#} & \mc{1}{c}{MSE} & \mc{1}{c}{MAE} & \mc{1}{c}{RMSE} & \mc{1}{c}{SDE} \\
\hline
& & \mc{4}{l}{All safe excitations} \\
CCSDTQ            &  41 & +0.03 & 0.03 & 0.04 & 0.03 \\
CC4               &  51 & +0.04 & 0.05 & 0.07 & 0.06 \\
CCSDT             &  63 & +0.28 & 0.28 & 0.35 & 0.20 \\
CC3               &  66 & +0.55 & 0.56 & 0.71 & 0.44 \\
SA-CASSCF         &  66 & +0.45 & 0.48 & 0.64 & 0.45 \\
CASPT2            &  66 & -0.16 & 0.23 & 0.31 & 0.26 \\
CASPT2(IPEA)      &  66 & +0.08 & 0.12 & 0.14 & 0.12 \\
CASPT3(IPEA)      &  66 & +0.11 & 0.13 & 0.18 & 0.14 \\
CASPT3            &  66 & +0.07 & 0.10 & 0.14 & 0.12 \\
SC-NEVPT2         &  66 & +0.13 & 0.14 & 0.19 & 0.14 \\
PC-NEVPT2         &  66 & +0.10 & 0.12 & 0.20 & 0.17 \\
\hline
& & \mc{4}{l}{Genuine doubly-excited states} \\
CCSDTQ            &  30 & +0.04 & 0.04 & 0.05 & 0.03 \\
CC4               &  31 & +0.07 & 0.07 & 0.09 & 0.06 \\
CCSDT             &  36 & +0.42 & 0.42 & 0.45 & 0.13 \\
CC3               &  36 & +0.91 & 0.91 & 0.95 & 0.25 \\
SA-CASSCF         &  36 & +0.41 & 0.46 & 0.65 & 0.51 \\
CASPT2            &  36 & -0.01 & 0.14 & 0.19 & 0.19 \\
CASPT2(IPEA)      &  36 & +0.05 & 0.11 & 0.13 & 0.12 \\
CASPT3(IPEA)      &  36 & +0.06 & 0.08 & 0.12 & 0.11 \\
CASPT3            &  36 & +0.03 & 0.08 & 0.12 & 0.11 \\
SC-NEVPT2         &  36 & +0.07 & 0.09 & 0.11 & 0.08 \\
PC-NEVPT2         &  36 & +0.07 & 0.08 & 0.18 & 0.17 \\
\hline
& & \mc{4}{l}{Partial doubly-excited states} \\
CCSDTQ            &  11 & +0.00 & 0.00 & 0.00 & 0.00 \\
CC4               &  20 & +0.00 & 0.01 & 0.01 & 0.01 \\
CCSDT             &  27 & +0.09 & 0.09 & 0.11 & 0.07 \\
CC3               &  30 & +0.13 & 0.13 & 0.15 & 0.07 \\
SA-CASSCF         &  30 & +0.51 & 0.51 & 0.63 & 0.37 \\
CASPT2            &  30 & -0.34 & 0.34 & 0.40 & 0.21 \\
CASPT2(IPEA)      &  30 & +0.10 & 0.13 & 0.15 & 0.11 \\
CASPT3(IPEA)      &  30 & +0.18 & 0.19 & 0.23 & 0.13 \\
CASPT3            &  30 & +0.11 & 0.14 & 0.16 & 0.11 \\
SC-NEVPT2         &  30 & +0.21 & 0.21 & 0.26 & 0.16 \\
PC-NEVPT2         &  30 & +0.14 & 0.17 & 0.22 & 0.17 \\
\end{tabular}
\end{ruledtabular}
\end{table}

In all cases, the mean absolute errors (MAEs) decrease as we move towards higher-order CC, which was expected.
Similarly, accounting for higher-order excitations, crucial for properly correlating doubly-excited states,
systematically decreases the mean signed errors (MSEs).
They always remain positive though, reflecting the well-known bias of CC towards the ground state.

\begin{figure*}
\includegraphics[width=\linewidth]{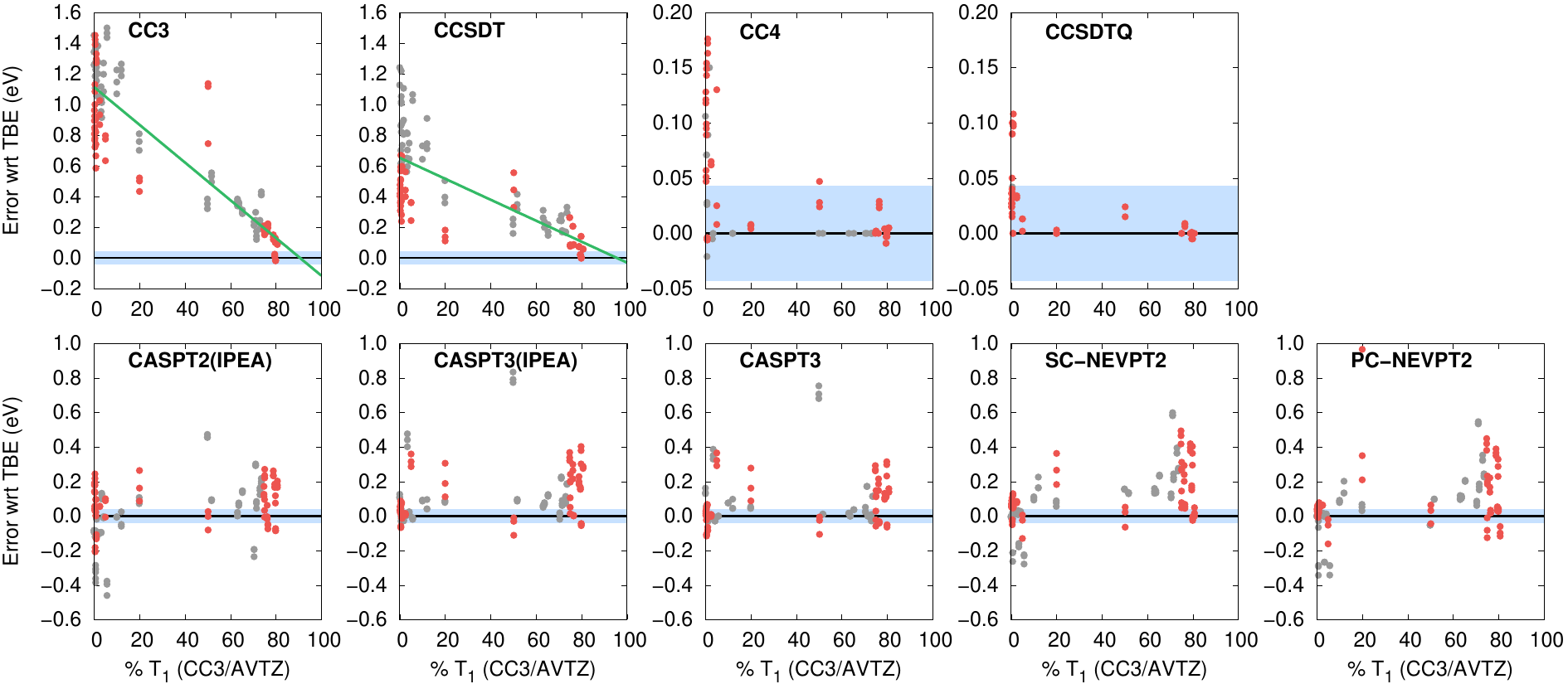}
\caption{Errors with respect to the TBEs, for various CC (top) and multiconfigurational (bottom) methods, as functions of the $\%T_1$ value obtained from CC3/AVTZ calculations,
for both safe (red) and unsafe (gray) states.
A linear fit to both safe and unsafe data points is shown as a green line, whereas chemical accuracy (\SI{0.043}{\eV}) is represented by the blue region.
\fk{Notice that different energy scales are used.}}
\label{fig:t1}
\end{figure*}

We find that partial double excitations are much better described than genuine doubles, across all levels of CC, by a factor of 5 to 8.
To further investigate this strong dependence on the character of the excited state,
we plot in Fig.~\ref{fig:t1} the errors with respect to the TBEs (accounting for the three basis sets) as a function of the $\%T_1$ value obtained from CC3/AVTZ. For the vast majority of states,
the value of $\%T_1$ usually depends very little on the choice of basis set.
It is clear that CC3 and CCSDT yield progressively smaller errors as $\%T_1$ increases, i.e., as the excited state assumes an increasingly singly-excited and less doubly-excited character.
Despite the large dispersion of errors in the regime of genuine doubles (small $\%T_1$), an overall linear trend is apparent. We will come back to this point in Sec.~\ref{sec:correction}.
Similarly, CC4 and CCSDTQ perform better for the partial than for the genuine doubles.
However, due to the fewer data points, and since these methods are often used to define the TBEs, it is not currently possible to infer how the errors evolve between the two groups of states.

Looking at how the errors behave along the CC hierarchy, starting with the partial doubles,
we find a relatively small MAE of \SI{0.13}{\eV} at the CC3 level, which decreases in CCSDT to \SI{0.09}{\eV}.
At the CC4 level, the errors are significantly reduced and reach chemical accuracy (MAE of only \SI{0.01}{\eV}).
CCSDTQ yields an even lower MAE, of \SI{0.003}{\eV}.
Excluding the partial doubles where CCSDTQ is used to generate the TBE increases its MAE to \SI{0.006}{\eV},
while doing the same for CC4 does not change its MAE.

Turning to the genuine doubly-excited states, CC3 only provides a qualitative description, and a large MAE of \SI{0.91}{\eV} is obtained. This is in sharp
contrast with the very small MAE (around \SI{0.03}{\eV}) obtained for states having a strong single-excitation character. \cite{Veril_2021}
Moving to CCSDT decreases the MAE by a factor of two, down to \SI{0.42}{\eV}.
The most significant improvement appears at the CC4 level, with a 6-fold smaller MAE of \SI{0.07}{\eV}.
Some improvement is observed with CCSDTQ, which yields a MAE of \SI{0.04}{\eV}, rendering this method barely chemically accurate for genuine doubles.
(Excluding the states where CCSDTQ is used to generate the TBE does not change its MAE.)
For a given pair of successive methods in the CC hierarchy, we observe a similar level of improvement for both partial and genuine doubles.

As for the multiconfigurational methods, we find more consistent results between partial and double excitations,
in comparison to more evident variations encountered in the CC methods.
There are, however, statistically meaningful differences.
As can be seen in Fig.~\ref{fig:t1},
the genuine doubles are more accurately described than the partial doubles, across all the multireference methods, contrasting with the trend observed for the CC methods.
The MAEs range from \SI{0.08}{\eV} to \SI{0.11}{\eV} for the genuine doubles, and from \SI{0.13}{\eV} to \SI{0.21}{\eV} for the partial doubles (excluding the less accurate CASPT2 method without an IPEA shift).
In addition, all multiconfigurational methods tend to overestimate the energies of partial doubly-excited states, with MSEs ranging from \SI{0.10}{\eV} to \SI{0.21}{\eV},
while for the genuine doubles the MSEs are overall closer to zero, ranging from \SI{0.03}{\eV} to \SI{0.07}{\eV} (again, excluding CASPT2 without IPEA shift).

Comparing all the multiconfigurational methods, CASTP2(IPEA) and CASPT3 are the most accurate ones for the partial doubles, with respective MAEs of \SI{0.13}{\eV} and \SI{0.14}{\eV}, comparable to that of CC3 (\SI{0.13}{\eV}),
though larger than that of CCSDT (\SI{0.09}{\eV}).
CASPT3 was also found to outperform the other multiconfigurational methods for singly-excited states. \cite{Boggio-Pasqua_2022}
As for the genuine doubles, we find that all multiconfigurational methods, except for CASPT2 and CASPT2(IPEA), yield surprisingly low and comparable MAEs, around \SI{0.08}{\eV}, virtually the same as obtained with CC4.

Applying an IPEA shift is a common practice in CASPT2 calculations, as it tends to improve the computed excitation energies. \cite{Zobel_2017,Sarkar_2022,Boggio-Pasqua_2022}
Here, we find that the IPEA shift has a much more pronounced effect for the partial doubles, where the MAE plummets from \SI{0.34}{\eV} to \SI{0.13}{\eV},
than for the genuine doubles, where the reduction is less impressive, from \SI{0.14}{\eV} to \SI{0.11}{\eV}.
Singly-excited states appear in-between, with respective MAEs of \SI{0.27}{\eV} and \SI{0.11}{\eV} according to comprehensive benchmarks reported in Refs.~\onlinecite{Sarkar_2022,Boggio-Pasqua_2022}.

In contrast to CASPT2, an IPEA shift does not change the overall performance of CASPT3 in the case of genuine doubles (MAEs of \SI{0.08}{\eV}),
and is detrimental in the case of partial doubles (MAE increases from \SI{0.14}{\eV} to \SI{0.19}{\eV}).
A slight worsening effect had also been observed for singly-excited states, with MAEs oscillating from \SI{0.09}{\eV} to \SI{0.11}{\eV}. \cite{Boggio-Pasqua_2022}
In light of this collection of results, applying an IPEA shift in CASPT3 calculations is probably not advisable.
While this represents an interesting advantage of CASPT3 over CASPT2(IPEA), there is no improvement for partial doubles (the MAE oscillates from \SI{0.13}{\eV} to \SI{0.14}{\eV}),
and a modest improvement for genuine doubles (the MAE decreases from \SI{0.11}{\eV} to \SI{0.08}{\eV}).
For singly-excited states, going from CASPT2(IPEA) to CASPT3 also has a marginal effect on the MAE, which decreases from \SI{0.11}{\eV} to \SI{0.09}{\eV}. \cite{Boggio-Pasqua_2022}

As for the two NEVPT2 approaches,
we find that PC-NEVPT2 is somewhat more accurate than SC-NEVPT2 for the partial doubles (MAEs of \SI{0.17}{\eV} and \SI{0.21}{\eV}, respectively),
whereas both schemes are equally accurate for genuine doubles, with MAEs of \SI{0.08}{\eV} and \SI{0.09}{\eV}.
A small preference in favor of PC-NEVPT2 was also pointed for singly-excited states, with respective MAEs of \SI{0.13}{\eV} and \SI{0.15}{\eV}. \cite{Sarkar_2022}

We also gauge how the basis set effect computed with a given method compares to that obtained with our TBEs.
The results for all the methods considered here, and for the three basis set differences, AVDZ $-$ Pop, AVTZ $-$ AVDZ, and AVTZ $-$ Pop,
are presented in the {\SupInf}.
The genuine doubly-excited state of ethylene has the most pronounced basis set effects, with a difference of \SI{0.49}{\eV} between TBE/AVTZ and TBE/Pop.
For this reason, this state is excluded in the calculation of average errors regarding the basis set correction.

The errors associated with CC3 have an overall positive sign in the regime of genuine double excitations.
This means that the CC3/AVDZ $-$ CC3/Pop difference is smaller than the TBE/AVDZ $-$ TBE/Pop reference value,
such that CC3 systematically underestimates the basis set effect in the case of genuine doubles.
Using CC3 to evaluate the basis set correction would introduce typical errors of around \SI{0.10}{\eV} for the AVDZ $-$ Pop basis set correction, \SI{0.05}{\eV} for the AVTZ $-$ AVDZ difference,
and a large \SI{0.15}{\eV} error for the AVTZ $-$ Pop difference.
This explains some of the main discrepancies, mentioned in Sec.~\ref{sec:Improvement} and~\ref{sec:Extension}, regarding the present and previous TBEs for genuine doubles.
The present findings clearly demonstrate that basis set corrections performed at the CC3 level introduce too large errors in the case of genuine double excitations.

This issue persists in CCSDT.
Somewhat surprisingly, the CCSDT average error for the AVTZ $-$ AVDZ difference is greater (\SI{0.08}{\eV}) than the corresponding CC3 average error (\SI{0.05}{\eV}).
Neither CC3 nor CCSDT are therefore recommended to perform a basis set correction for genuine doubly-excited states.
Conversely, CC4 and CCSDTQ can be safely employed, as they introduce errors below \SI{0.02}{\eV} and \SI{0.01}{\eV}, respectively.
The genuine doubly-excited state of formaldehyde is a notable exception, though, where CC4 yields an absolute error of around \SI{0.1}{\eV} for the AVTZ $-$ AVDZ and AVDZ $-$ Pop basis set corrections.
Even though this large error does not pose a problem in defining our TBEs for formaldehyde (which rely on exFCI results),
it highlights that even CC4 is not always reliable for the basis set correction.

The performance of multiconfigurational methods is \fk{somewhat worse than} that of CC4,
with CASPT3 delivering the smallest basis set correction errors, between \SI{0.02}{\eV} and \SI{0.04}{\eV}.
Comparing different methods for the basis set corrections can thus be useful to identify the consistency of the correction or the presence of a more problematic state.

In contrast to the case of genuine doubles, the basis set correction obtained with CC methods is much more well-behaved for the partial doubly-excited states.
CC3 tends to slightly underestimate the basis set effect.
The absolute errors lie within the region of chemical accuracy, averaging at \SI{0.01}{\eV} for the AVTZ $-$ AVDZ and AVDZ $-$ Pop differences and at \SI{0.02}{\eV} for the AVTZ $-$ Pop difference.
CCSDT and higher levels of CC are very safe for the basis set correction of partial doubles, with errors below \SI{0.01}{\eV}.
Multiconfiguration methods have errors ranging from \SI{0.02}{\eV} to \SI{0.05}{\eV} and should therefore be avoided if CC3 results are available.
We underline that the average errors for multiconfigurational methods have opposite signs in the AVDZ $-$ Pop and AVTZ $-$ AVDZ basis set effects,
which partially cancels out in the AVTZ $-$ Pop difference, which ends up having smaller absolute errors.

Despite being small in absolute terms, the errors stemming from the basis set correction should be considered when evaluating the TBEs.
If chemical accuracy is desired, the calculation performed with the smaller basis should be more stringent than usual, depending on the typical errors associated with the basis set correction.

\fk{For four systems, we have compared the present TBEs with available DMC results, which can also be considered a reference method.
For nitroxyl, the two approaches perfectly agree, while for the $^1A_g$ state of tetrazine, the difference is small (\SI{-0.04}{\eV}).
Our TBEs appear more redshifted with respect to DMC
for glyoxal (\SI{-0.14}{\eV}) and for the $^1A_1$ genuine/partial double of cyclopentadienone (\SI{-0.11}{\eV}/\SI{-0.18}{\eV}).
It is not clear which approach renders excitation energies closer to the exact values.
Our protocol for obtaining the TBEs is state-dependent, as discussed above for each case, but the associated uncertainties are probably smaller than the observed differences to DMC.
Meanwhile, the fixed-node error associated with DMC calculations is hard to estimate.
In addition, our TBEs are given for the aug-cc-pVTZ basis set, whereas DMC is a real-space method and thus corresponds to results closer to the complete basis set limit.
However, further augmenting the basis set is expected to slightly lower the excitation energies and increase the gap between the two approaches.}

\section{Correction based on $\%T_1$}
\label{sec:correction}

We mentioned before that a linear trend (see Fig.~\ref{fig:t1}) can be observed for the errors of CC3 and CCSDT when plotted as functions of the $\%T_1$ value (computed at the CC3/AVTZ level).
Based on this observation, we propose a simple additive correction to the computed excitation energies, denoted as +LT1 (which stands for linear $\%T_1$).
When applied to CC3, for instance, it gives rise to the CC3+LT1 model.
By fitting our full set of data points (safe and unsafe results, and the three basis sets) to the linear function $a \times \%T_1/100 + b$ (the green lines in Fig.~\ref{fig:t1}),
the computed excitation energy $\Delta E$ is shifted to $\Delta E (+\text{LT1}) = \Delta E - [ a \times \%T_1/100 + b]$.
The fitting parameters $a$ and $b$ are presented in the {\SupInf}.

This type of correction can be useful when the errors of a given method are relatively consistent, or, in other words, if its standard deviation of error (SDE) is smaller than its MAE.
This is evident for CC3, which yields systematically overestimated excitation energies for the genuine doubles,
with a dispersion in energy (SDE of \SI{0.25}{\eV}) smaller than the overall absolute errors (MAE of \SI{0.91}{\eV}) by a factor of around 3.6.
Similarly, such a correction can be helpful to correct the genuine doubles computed with CCSDT, which has a corresponding SDE (\SI{0.13}{\eV}) smaller than its MAE (\SI{0.42}{\eV}) by a factor of 3.2.
When applied to CC3 and CCSDT, the +LT1 correction is therefore a simple attempt to recover from the lack of higher-order excitations that would correlate the genuine doubly-excited states.
Because the fitting procedure accounts for both genuine and partial doubles, and the latter also has smaller SDEs than MAEs, both types of states are impacted by the correction.

\begin{figure}
\includegraphics[width=\linewidth]{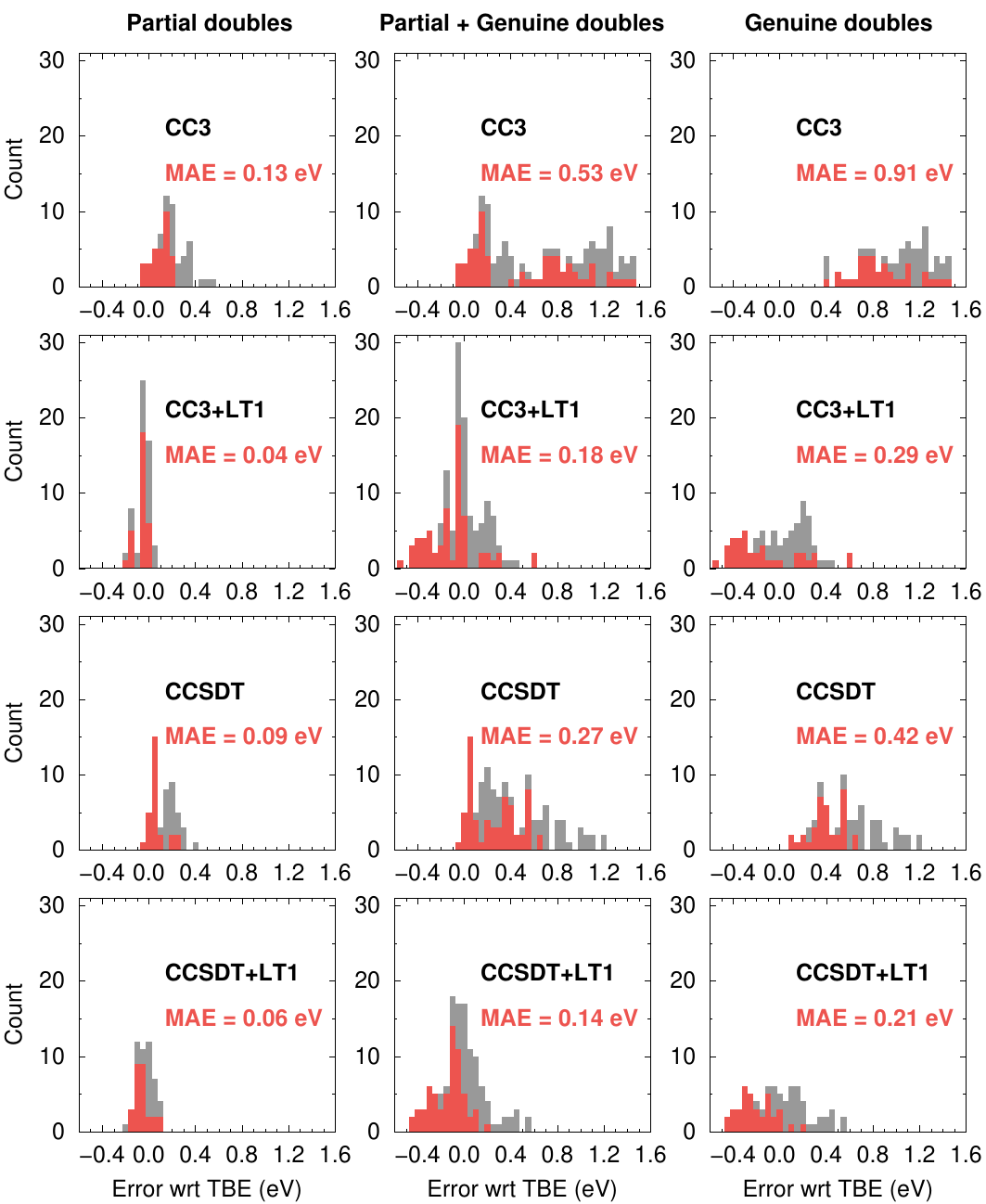}
\caption{Distribution of errors with respect to the TBEs (of all basis sets), computed at the CC3, CC3+LT1, CCSDT, and CCSDT+LT1 levels of theory (see text for explanation of the LT1 correction),
restricted to partial (left) or genuine (right) doubles, and for both partial and doubles (center), including safe (red) and both safe and unsafe (gray) states.}
\label{fig:lt1}
\end{figure}

In Fig.~\ref{fig:lt1}, we show the distribution of errors obtained with CC3, CCSDT, and their +LT1 corrected counterparts.
Safe states are represented in red and unsafe states in gray.
We find that the +LT1 correction significantly reduces the MAEs of CC3, by a factor ranging between 3 and 3.5. The relative improvement is comparable for both partial and genuine doubles.
From CC3 to CC3+LT1, the MAE for the partial doubles decreases from \SI{0.13}{\eV} to \SI{0.04}{\eV}, whereas, for the genuine doubles, it reduces from \SI{0.91}{\eV} down to \SI{0.29}{\eV}.
While these values concern the safe states, the same conclusions also hold when accounting for the unsafe states.
The observed difference on their underlying distribution of errors reflects the system size in each subset:
safe states comprise mostly smaller systems (thus smaller errors), whereas unsafe states include larger systems (and larger errors).

The conclusions obtained when comparing CCSDT and CCSDT+LT1 are similar to their CC3 analogs.
The main difference is that the LT1 correction reduces the errors of CCSDT by a factor ranging between 1.5 and 2, whereas in CC3, this factor is found to range between 3 and 3.5.
For the partial doubles, the MAE decreases from \SI{0.09}{\eV} to \SI{0.06}{\eV}, and for the genuine doubles they shift from \SI{0.42}{\eV} to \SI{0.21}{\eV}.
\fk{For partials, we find that CCSDT+LT1 and CC3+LT1 have comparable performances (MAE of \SI{0.06}{\eV} and \SI{0.04}{\eV}, respectively),
whereas for genuines the former method is somewhat more accurate (MAE of \SI{0.21}{\eV} and \SI{0.29}{\eV}).
Importantly,}
we find that CC3+LT1 outperforms CCSDT, \fk{for both partials and genuines,} despite being a cheaper approach.

Because there are fewer CC4 and CCSDTQ data points, especially in the intermediate $\%T_1$ region, and considering that these methods are often used to define the TBEs,
we cannot conclude that a linear trend would still be observed.
Despite that, the +LT1 correction (based on the available CC4 data) brings down the overestimated energies for the genuine doubles computed with CC4, reducing its MAE from \SI{0.07}{\eV} to \SI{0.05}{\eV}.
For the multiconfigurational methods, the correction can be applied to improve the energies of the partial double excitations, which appear generally overestimated (see Fig.~\ref{fig:t1}),
while they have virtually no effect on the genuine doubles.
The overall improvements are more modest than those observed for CC3 and CCSDT.
The MAE decreases from \SI{0.14}{\eV} in CASPT3 to \SI{0.09}{\eV} in CASPT3+LT1, for instance.
The full set of results are reported in the {\SupInf}.

We further recall that our fitting procedure accounts for safe and unsafe states.
In this way, CC3+LT1 and CCSDT+LT1 tend to underestimate the excitation energies of the safe ones, while overestimating the unsafe ones.
Had we included only the states whose TBEs are considered safe, their errors would certainly be smaller, at the expense of larger errors for the unsafe ones.
Naturally, one could generalize this simple +LT1 correction to also account for the system size, but we did not pursue this here.

\section{Conclusion}
\label{sec:conclusion}

We have performed a systematic investigation on the vertical excitation energies of molecular states that have strong doubly-excited character.
The present set comprises 28 genuine doubles, which are dominated by a doubly-excited character,
plus 19 partial doubles, where this character is not dominant but is still important.
Based on CIPSI, high-order CC, and multiconfigurational calculations, TBEs for this set of 47 states are reported.

Compared to our previous benchmark study on doubly-excited states, \cite{Loos_2019c}
the more substantial number of states considered here (47 against 20), especially on partial doubles,
in addition to the more reliable TBEs,
allow us to draw solid conclusions about the performance of different methodologies.
CC methods are found to be progressively more accurate as the doubly-excited character becomes less prominent, from the genuine doubles to the partial doubles and finally to the singly-excited states.
Multiconfigurational methods, in contrast, are similarly accurate for genuine doubles and singly-excited states, but less accurate for the partial doubles.

To obtain chemically accurate excitation energies of genuine doubles, one needs at least CCSDTQ (MAE of \SI{0.04}{\eV}),
whereas CC4 is already very accurate for the partial doubles (MAE of \SI{0.01}{\eV}).
If the desired accuracy is less stringent, CC4, CASPT3, or NEVPT2 are reasonable choices for genuine doubles, all having MAEs of around \SI{0.08}{\eV}.
For the partial doubles, CCSDT and CC3 are surprisingly accurate (MAE of \SI{0.09}{\eV} and \SI{0.13}{\eV}, respectively).
In contrast, multiconfigurational methods struggle more in these cases,
with the smaller MAEs provided by CASPT2(IPEA) (\SI{0.13}{\eV}) and CASPT3 (\SI{0.14}{\eV}).
\fk{While the present calculations were performed with a real level shift, it would be interesting to gauge the performance of CASPT2 and CASPT3 when employed with imaginary shifts or different regularization techniques. \cite{Battaglia_2022}}

CC3 is the lowest-order CC method providing a qualitatively correct description of doubly-excited states.
However, it yields systematically overestimated excitation energies.
We have demonstrated that the $\%T_1$ value computed with CC3 can reasonably predict the method's inaccuracy in describing doubly-excited states.
This motivated us to introduce a simple correction (labeled +LT1) that can be employed after a CC3 calculation, significantly improving estimates of excitation energies.
Such a correction reduces the MAEs obtained with CC3 by a factor of 3,
from \SI{0.91}{\eV} down to \SI{0.27}{\eV} for the genuine doubles,
and from \SI{0.13}{\eV} down to \SI{0.04}{\eV} for the partial doubles, whereas we do not recommend it for the singly-excited states.

To conclude, the comprehensive and accurate set of TBEs reported here helps in assessing the performance of known excited-state methods. Likewise, we hope they
will be helpful for gauging novel methodologies aiming to describe doubly-excited states.

\acknowledgements{
This work was performed using the HPC resources from CALMIP (Toulouse) under allocation 2024-18005, and of the GliCiD/CCIPL computational center installed in Nantes.
This project has received funding from the European Research Council (ERC) under the European Union's Horizon 2020 research and innovation programme (Grant agreement No.~863481).
}

\section*{Supporting information}

MAEs for the excitation energies,
MAEs for the basis set effect,
errors of the basis set effects as functions of $\%T_1$,
distribution of errors on the excitation energies,
specific orbitals involved in each electronic transition,
additional details about the multiconfigurational calculations,
raw data on the excitation energies,
and additional statistical results.


\section*{References}
\bibliography{doubles}

\end{document}